\documentclass[iop,revtex4]{emulateapj}
\slugcomment{{\sc Accepted to ApJ:} March 10, 2014}

\newcommand{\Lsun}{$L_{\odot}$}
\newcommand{\Msun}{$M_{\odot}$}
\newcommand{\Rsun}{$R_{\odot}$}

\shorttitle{Nature of deeply embedded OMC-2 FIR 4}
\shortauthors{Furlan et al.}

\begin{document}

\title{On the nature of the deeply embedded protostar OMC-2 FIR 4}

\author{E. Furlan\altaffilmark{1,2}, S. T. Megeath\altaffilmark{3},
M. Osorio\altaffilmark{4}, A. M. Stutz \altaffilmark{5},
W. J. Fischer\altaffilmark{3}, B. Ali\altaffilmark{6}, T. Stanke\altaffilmark{7}, \\
P. Manoj\altaffilmark{8}, J. D. Adams\altaffilmark{9}, and J. J. Tobin\altaffilmark{10}}
\altaffiltext{1}{National Optical Astronomy Observatory, 
950 N. Cherry Avenue, Tucson, AZ 85719, USA; furlan@ipac.caltech.edu}
\altaffiltext{2}{Visitor at the Infrared Processing and Analysis Center, California
Institute of Technology, 770 S. Wilson Ave., Pasadena, CA 91125, USA}
\altaffiltext{3}{Ritter Astrophysical Observatory, Department of Physics and
Astronomy, University of Toledo, 2801 W. Bancroft Street, Toledo, OH 43606, USA}
\altaffiltext{4}{Instituto de Astrof\'isica de Andaluc\'ia, CSIC, Camino Bajo de 
Hu\'etor 50, E-18008 Granada, Spain}
\altaffiltext{5}{Max-Planck-Institut f\"ur Astronomie, K\"onigstuhl 17, D-69117
Heidelberg, Germany}
\altaffiltext{6}{NHSC/IPAC, California Institute of Technology, 770 S. Wilson Ave., 
Pasadena, CA 91125, USA}
\altaffiltext{7}{ESO, Karl-Schwarzschild-Strasse 2, D-85748 Garching bei 
M\"unchen, Germany}
\altaffiltext{8}{Department of Astronomy and Astrophysics, Tata Institute
of Fundamental Research, Homi Bhabha Road, Colaba, Mumbai 400005, India}
\altaffiltext{9}{USRA - SOFIA, DAOF, 2825 E. Ave. P, Palmdale, CA 93550}
\altaffiltext{10}{{\it Hubble Fellow}; National Radio Astronomy Observatory,
Charlotttesville, VA 22903, USA}

\begin{abstract}

We use mid-infrared to submillimeter data from the {\it Spitzer}, 
{\it Herschel}, and APEX telescopes to study the bright sub-mm source 
OMC-2 FIR 4. We find a point source at 8, 24, and 70 $\mu$m, and a 
compact, but extended source at 160, 350, and 870 $\mu$m. The peak 
of the emission from 8 to 70 $\mu$m, attributed to the protostar associated 
with FIR 4, is displaced relative to the peak of the extended emission; the 
latter represents the large molecular core the protostar is embedded within. 
We determine that the protostar has a bolometric luminosity of 37 \Lsun, 
although including more extended emission surrounding the point source 
raises this value to 86 \Lsun. Radiative transfer models of the protostellar
system fit the observed SED well and yield a total luminosity of most likely 
less than 100 \Lsun. Our models suggest that the bolometric luminosity 
of the protostar could be just 12-14 \Lsun, while the luminosity of the 
colder ($\sim$ 20 K) extended core could be around 100 \Lsun, with a 
mass of about 27 \Msun. 
Our derived luminosities for the protostar OMC-2 FIR 4 are in direct 
contradiction with previous claims of a total luminosity of 1000 \Lsun\ 
\citep{crimier09}. Furthermore, we find evidence from far-infrared 
molecular spectra \citep{kama13, manoj13} and 3.6 cm emission 
\citep{reipurth99} that FIR 4 drives an outflow. The final stellar mass 
the protostar will ultimately achieve is uncertain due to its association 
with the large reservoir of mass found in the cold core. 

\end{abstract}


\section{Introduction}

The OMC 2 region in the Orion A star-forming complex is actively forming
low- and intermediate-mass stars \citep{peterson08}. It lies in the 
northern part of the extended Orion Nebula Cluster and is embedded in a
2 pc long, narrow filament extending away from the Orion Nebula itself
\citep[][Megeath et al., in preparation]{chini97,carpenter00}. OMC 2 contains some 
of the most luminous infrared and sub-mm sources in the Orion A molecular 
cloud outside of the Orion Nebula \citep{johnson90,mezger90}. Over the last 
few decades, several surveys from infrared to radio wavelengths disentangled 
the multitudes of sources found in this region, revealing young stellar objects in 
different evolutionary stages, ranging from deeply embedded protostars to young 
stars surrounded by disks 
\citep{gatley74, rayner89, johnson90, mezger90, jones94, ali95, chini97, 
lis98, reipurth99, nielbock03, tsujimoto03, peterson08, megeath12, adams12}.

The first near-IR images of OMC 2 by \citet{gatley74} revealed a small cluster of
five bright IR sources in a region 90\arcsec, or 0.2 pc, in diameter. These have
subsequently been shown to be young stellar objects with luminosities ranging from
20 to 300~\Lsun\ \citep{adams12}. Subsequent sub-mm and mm imaging 
\citep{mezger90,chini97,lis98} showed that in the center of this small
cluster is a bright sub-mm source. This object, OMC-2 FIR 4, is the brightest
sub-mm (350-1300~$\mu$m) source the OMC 2 region. It is connected through
filamentary structures to two other adjacent sources that are bright at sub-mm
wavelengths and are coincident with two of the bright IR sources of \citet{gatley74}:
OMC-2 FIR 3 matches a protostar $\sim$ 28\arcsec\ to the north (also known 
as SOF 2N or HOPS 370), while OMC-2 FIR 5 agrees with a protostar ~$\sim$ 
17\arcsec\ to the south (SOF 4 or HOPS 369, see \citealt{adams12}). 
Outside of the massive star-forming region OMC-1 in the Orion Nebula, FIR 4 
is the brightest 870~$\mu$m source in the Orion A cloud (Stanke et al. 2014, 
in preparation).  Although bright in the sub-mm, FIR 4 was not detected in 
the near-IR by \citet{tsujimoto03} and only tentatively associated with a 
near- to mid-IR source by \citet{nielbock03}. The detection of a 3.6 cm 
source with the VLA  toward FIR 4 was the first compelling evidence that the 
sub-mm source contained a deeply embedded protostar; the elongated radio 
source was interpreted as free-free emission originating from shock-ionized gas 
in an outflow launched by a protostar \citep{reipurth99}.

FIR 4 also coincides with the IRAS source 05329-0512. Its bolometric luminosity,
integrated over an area of 50\arcsec$\times$50\arcsec\ around it, 
was estimated to be 420 \Lsun\ \citep{mezger90}. FIR 4 was thus identified 
and studied as an intermediate-mass protostar \citep{johnstone03, crimier09}. 
\citet{crimier09} constructed a spectral energy distribution (SED) for FIR 4 by 
retrieving archived mid-infrared to millimeter observations and extracting fluxes. 
They modeled the SED and derived a total luminosity of 1000 \Lsun. 
More recently, the infrared emission from a protostar toward FIR 4 (known as 
SOF 3 or HOPS 108) was resolved by \citet{adams12} using 2\arcsec\ to 
19\arcsec\ resolution data from the {\it Spitzer Space Telescope} \citep{werner04},
the {\it Stratospheric Observatory For Infrared Astronomy} \citep[SOFIA;][]{young12}, 
the {\it Herschel Space Telescope}\footnote{{\it Herschel} is an ESA space 
observatory with science instruments provided by European-led Principal Investigator 
consortia and with important participation from NASA.} \citep{pilbratt10}, and from 
the Atacama Pathfinder Experiment (APEX) telescope. This work has cast doubt on 
the high luminosity of OMC-2 FIR 4; modeling of the SEDs by \citet{adams12} 
found that the intrinsic luminosity lies in the 30-50 \Lsun\ range. These data 
also showed that within the beam of IRAS, the other near-IR sources originally 
found by \citet{gatley74} dominated the flux out to 70 $\mu$m with 
luminosities varying from 20 to 300~\Lsun; OMC-2 FIR 3 (SOF 2N, HOPS 370) 
was found to be the most luminous source in the region.  

Only at wavelengths $\gtrsim$ 160~$\mu$m does FIR 4 dominate; however, 
it is unclear whether the entire sub-mm emission is associated with the protostar 
observed at shorter wavelengths. Millimeter interferometry by \citet{shimajiri08} 
resolved FIR 4 into 11 dusty cores. Furthermore, they found that high-velocity 
gas traced by CO is dominated by an outflow from FIR 3.  
They proposed that the motion seen in the dense gas toward FIR 4 could be 
explained by the interaction of the powerful outflow from FIR 3 with the FIR 4 
clump. On the basis of interferometric observations made in both continuum 
and line, \citet{lopez13} interpreted FIR 4 as containing three distant components,  
a western core, a southern core and a main core containing a young star, with a 
total mass of 9.2 to 25.7~\Msun.  Noting the lack of the detection of outflow 
signatures from FIR 4 in their interferometric observations and the proposal of 
\citet{shimajiri08} that motions in FIR 4 are driven by an outflow from FIR 3,
they suggested that the 3.6 cm source is due to photo-ionization of gas by  
an early-type (B3$-$B4) star with a luminosity of 700-1000 \Lsun\ within one 
of the three components.  

In this publication, we use {\it Spitzer}, {\it Herschel}, and APEX imaging 
of FIR 4 from 3.6 to 870~$\mu$m obtained for the Herschel Orion Protostar 
Survey (HOPS) to study the protostar associated with FIR 4, with the goal of 
resolving the large uncertainties in the luminosity of the protostar and its relationship 
to the sub-mm clump. We use these data to measure the SED of the protostar  
and constrain its bolometric luminosity and temperature, exploring the effect of the 
choice of aperture size, or the use of PSF fitting photometry, on the final result.  
By using radiative transfer models, we explore the range of possible luminosities and
source properties and show that a wide range of luminosities are possible. We also
investigate the relationship between the protostar and sub-mm clump and the
possibility that much of the sub-mm luminosity is due to external heating. We favor 
a model that has a deeply embedded, protostar with L $<$ 100~\Lsun\ driving 
an outflow, forming on the side of a massive ($\sim$ 30~\Msun) clump. 


\section{Data Overview}
\label{data}

\begin{figure*}[!]
\centering
\includegraphics[scale=0.75]{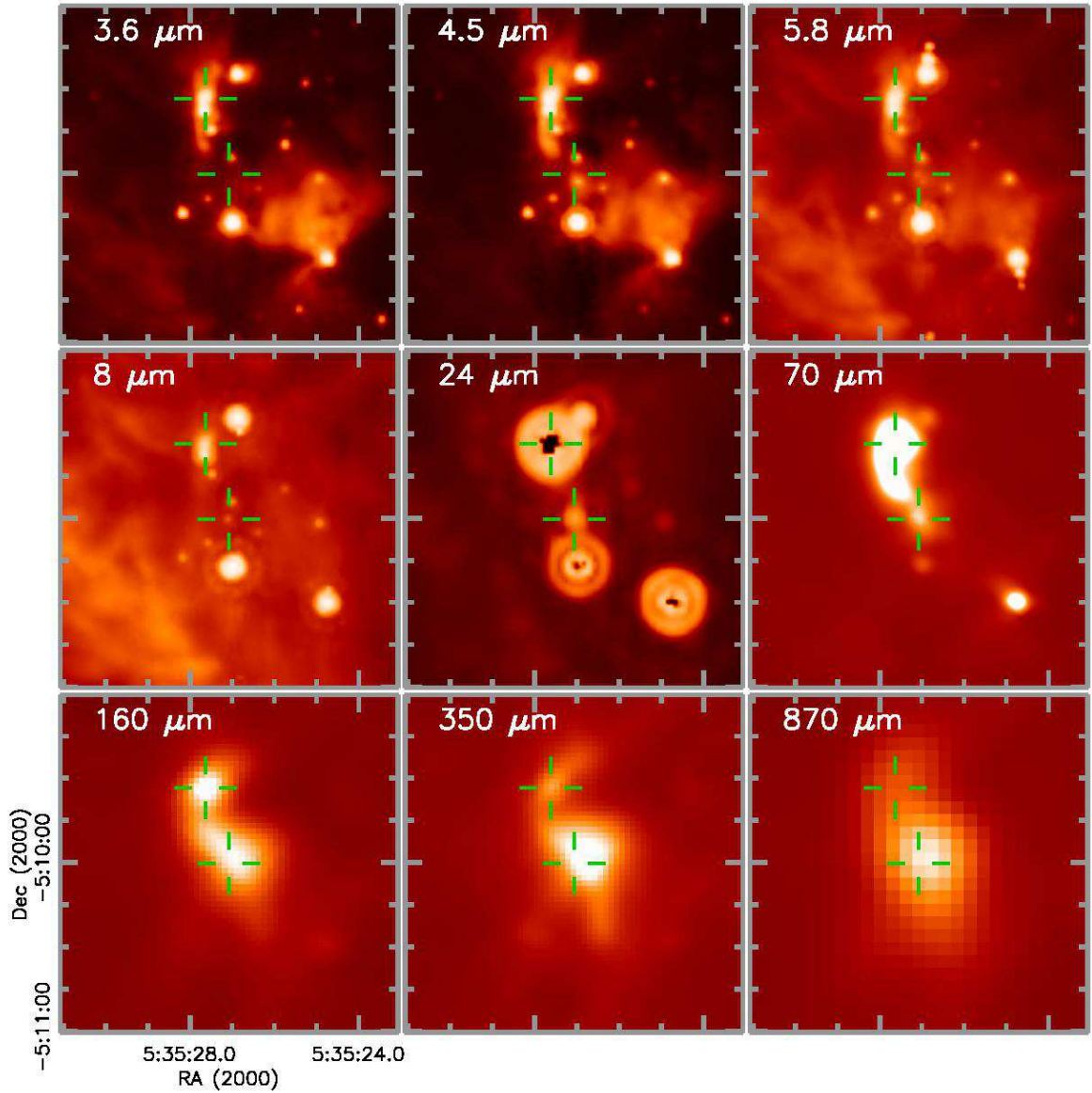}
\caption{IRAC 3.6, 4.5, 5.8, and 8.0 $\mu$m, MIPS 24 $\mu$m,
PACS 70 and 160 $\mu$m, SABOCA 350 $\mu$m and LABOCA
870 $\mu$m images of the region around OMC-2 FIR 4 (SOF 3, HOPS 108).
At 24 $\mu$m, the saturated sources to the north, south, and southwest
are OMC-2 FIR 3 (SOF 2N, HOPS 370), OMC-2 FIR 5 (SOF 4, HOPS 369), 
and SOF 5 (HOPS 368), respectively. The two crosshairs show the position of
FIR 3 (northern crosshairs) and FIR 4 (southern crosshairs).
\label{HOPS108_images}}
\end{figure*}

In this section, we present our data on OMC-2 FIR 4 and explain how
we extracted the photometry in the far-IR and sub-mm.
OMC-2 FIR 4 was detected by the {\it Spitzer} InfraRed Array Camera 
\citep[IRAC;][]{fazio04} and Multiband Imaging Photometer for Spitzer 
\citep[MIPS;][]{rieke04} at 8.0 and 24 $\mu$m, respectively 
(\citealt{megeath12}; see Figure~\ref{HOPS108_images}). 
A mid-infrared spectrum (5-37 $\mu$m) was also taken using the InfraRed 
Spectrograph \citep[IRS;][]{houck04} on {\it Spitzer}. 
FIR 4 was also detected at 37.1 $\mu$m using FORCAST \citep{herter12} 
on SOFIA (source SOF 3 of \citealt{adams12}). 
As part of the Herschel Orion Protostar Survey (HOPS), a {\it Herschel} 
open-time key program \citep[e.g.,][Fischer et al. 2014, in preparation; 
Ali et al. 2014, in preparation] {fischer10, stanke10,stutz13,manoj13}, 
it was observed with the Photodetector Array Camera and Spectrometer 
\citep[PACS;][]{poglitsch10} at 70 and 160 $\mu$m (see 
Figure~\ref{HOPS108_images}). In the HOPS catalog, OMC-2 FIR 4 is 
source HOPS 108. It was also observed with PACS at 100 $\mu$m 
by the Gould Belt Survey \citep[e.g.,][]{andre10}. 
In the submillimeter, OMC 2 was mapped at 350 and 870 $\mu$m with the 
SABOCA and LABOCA instruments \citep[][respectively]{siringo10, siringo09} 
on the APEX telescope (see Figure \ref{HOPS108_images}).
Table \ref{obs_data} displays the fluxes extracted from these data sets, 
as well as some measurements from the literature. Details on the data reduction
and photometry for the PACS data can be found in Ali et al. (2014, in preparation),
while details on the measurements of the APEX fluxes can be found in
\citet{stutz13}, Stutz et al. (2014, in preparation), and Stanke et al.
(2014, in preparation).

\begin{deluxetable*}{cccc}
\tabletypesize{\footnotesize}  
\tablewidth{0pt}
\tablecaption{{Photometry of OMC-2 FIR 4}
\label{obs_data}}
\tablehead{
\colhead{Wavelength} & \colhead{Flux [Jy]} & 
\colhead{Aperture radius} & \colhead{Reference}}
\startdata
{\bf 8 $\mu$m} & {\bf 0.03} & 2.4\arcsec & \citet{megeath12} \\
{\bf 24 $\mu$m} & {\bf 1.519} & PSF & \citet{megeath12} \\
37.1 $\mu$m & 8.4 & 4.3\arcsec\ beam & \citet{adams12} \\
{\bf 70 $\mu$m} & {\bf 132.5} & 9.6\arcsec & this work \\
{\bf 70 $\mu$m} & {\bf 40.81} & PSF & this work \\
{\bf 100 $\mu$m}  & {\bf 287.7} & 9.6\arcsec & this work  \\
{\bf 160 $\mu$m}  & {\bf 611.2} & 12.8\arcsec & this work \\
{\bf 160 $\mu$m} & {\bf 270.2} & PSF & this work \\
350 $\mu$m  & 67 & 12\arcsec\ beam & \citet{lis98} \\
{\bf 350 $\mu$m}  & {\bf 54.7} & 7.34\arcsec & this work \\
{\bf 350 $\mu$m} & {\bf 43.2} & 7.34\arcsec\ beam & this work \\
850 $\mu$m  & 7.5 & 14\arcsec\ beam & \citet{johnstone99} \\
{\bf 870 $\mu$m}  & {\bf 12.3} & 19\arcsec & this work \\
{\bf 870 $\mu$m} & {\bf 8.39} & 19\arcsec\ beam & this work\\
1.3 mm & 8.0  & 50\arcsec\ $\times$ 50\arcsec & \citet{mezger90} \\
1.3 mm & 1.252 & 22\arcsec\ $\times$ 17\arcsec & \citet{chini97} \\
2.0 mm & 1.06 & 4.87\arcsec\ $\times$ 2.73\arcsec & \citet{lopez13} \\
3.6 cm & 6.4 $\times 10^{-4}$  & $\sim$ 8\arcsec\ beam & \citet{reipurth99}
\enddata
\tablecomments{The flux values we use for our SED models are shown in
bold (see text for details). In the column labeled ``aperture radius'', ``PSF''
means that the flux was determined via PSF photometry, and a numerical
value followed by ``beam'' indicates that the flux is the peak beam flux of
the source as measured in a beam with the specified FWHM.}
\end{deluxetable*}

In the {\it Spitzer} IRAC images, OMC-2 FIR 4 is faint, but clearly detected, 
at 8.0 $\mu$m. Some emission can also be seen at 4.5 $\mu$m and 
5.8 $\mu$m, but there is no well-defined point source at these wavelengths.
At 5.8 $\mu$m, there seem to be two emission peaks, one that matches
the 8 and 24 $\mu$m position, and one slightly offset, while at 4.5 $\mu$m 
there is a strong emission peak only at the offset position. The detection of 
emission offset relative to the peak position seen at 8-70 $\mu$m could be 
an indication of an outflow (see section \ref{outflow}). 
About 6\arcsec\ to the north of FIR 4 lies an object that is brighter in all IRAC 
bands, but much fainter at 24 $\mu$m and not detected at 70 $\mu$m 
and longer wavelengths (see Figure \ref{HOPS108_images}). This is source 
MIR 24 tentatively identified with FIR 4 by \citet{nielbock03}, but it is a
separate source (also known as HOPS 64 in the HOPS catalog).

The IRS spectrum of FIR 4 is very noisy in the 5-14 $\mu$m region, mostly 
due to deep ice and silicate absorption features, but at 8 $\mu$m agrees 
with the IRAC measurement within $\sim$20\%. Given the slit widths of 3.6\arcsec\ 
for the Short-Low module (SL; 5-14 $\mu$m) and 10.5\arcsec\ for the 
Long-Low module (LL; 14-37 $\mu$m), as well as the slit orientations, none 
of the bright neighboring sources contaminated the IRS spectrum. Only HOPS 64, 
located 6.3\arcsec\ to the north of FIR 4, partially entered the LL slit, but its 
flux contribution at wavelengths $\gtrsim$ 15 $\mu$m is small (its MIPS 24 
$\mu$m flux is 0.57 Jy, compared to 1.5 Jy for FIR 4). There is a discrepancy 
between the MIPS 24 $\mu$m flux of FIR 4 and the 24 $\mu$m flux derived 
from IRS spectra in that the IRS flux is a factor of 2 too high. This could be due to 
the fact that more extended emission from the filament was included in the IRS 
measurement (slit width of 10.5\arcsec\ compared to the typical FWHM of 
the MIPS 24 $\mu$m  PSF of $\sim$ 6\arcsec). The IRS spectrum 
was scaled by 0.5 to match the MIPS 24 $\mu$m flux. 
When compared to the SOFIA/FORCAST measurement at 37.1 $\mu$m from
\citet{adams12}, where FIR 4 appears as a point source, the IRS spectrum
is about a factor of 1.3-1.9 too high (the range considers the calibration 
uncertainty of the 37.1 $\mu$m flux), roughly consistent with the discrepancy 
found for the MIPS 24 $\mu$m flux.  

\begin{figure}[!t]
\centering
\includegraphics[scale=0.54]{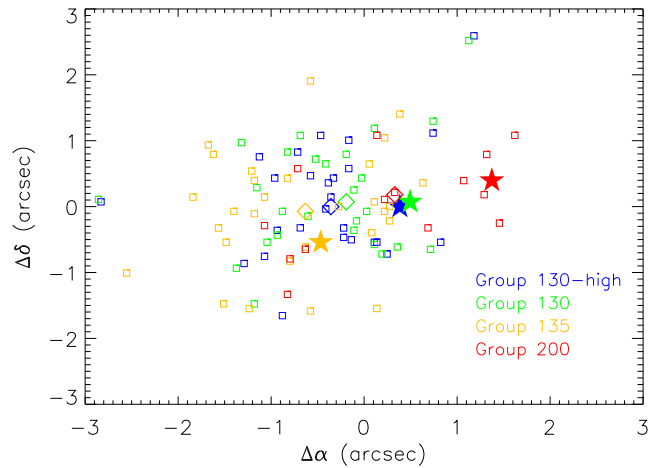}
\caption{Offsets in right ascension and declination relative to the 
{\it Spitzer} positions for all PACS 70 $\mu$m sources with {\it Spitzer}
analogs in the fields that contain the source OMC-2 FIR 4. The targets
were observed by {\it Herschel} in four different scan groups, represented
by four colors. The squares identify all HOPS targets observed in these groups, 
with FIR 4 shown with a filled star symbol. The open diamonds represent the 
median offsets found for the objects in the four groups.
\label{HOPS108_P70_positions}}
\end{figure}

At 70 and 160 $\mu$m, emission towards OMC-2 FIR 4 can be clearly discerned 
(Figure \ref{HOPS108_images}). FIR 4 is part of a dense filament extending from 
FIR 3 (HOPS 370) to the north. At 70 $\mu$m, a point source can be seen 
near the position of the Spitzer 8 and 24 $\mu$m source. To compare the
position of the PACS sources to those in the {\it Spitzer} images, the PACS
maps have been re-centered based on the average offsets between the
{\it Spitzer} and PACS 70 $\mu$m observations of the HOPS targets in 
the field (see Figure \ref{HOPS108_P70_positions}). The offsets were
determined independently for the four distinct images constructed from the
four separate groups of PACS scans that covered FIR 4 as part of the HOPS
program. In these four images, the 70 $\mu$m position of FIR 4 is offset 
from the {\it Spitzer} position by 0.4\arcsec\ to 1.4\arcsec\ (Figure 
\ref{HOPS108_P70_positions}), much smaller than the FWHM of 7\arcsec\ 
at 70 $\mu$m. These offsets are comparable to the offsets found for the 
other HOPS sources in each group and match the positional uncertainty 
expected from the {\it Herschel} pointing accuracy of $\sim$ 2\arcsec. 
The offset between the {\it Spitzer} and PACS 70 $\mu$m data in right 
ascension is $\sim$ 0.7\arcsec-1\arcsec\ larger for FIR 4 than the 
median offset for the other 70 $\mu$m sources in three of the four
images; for the fourth image (constructed from group 135 scans), the offset for 
FIR 4 and the median offset agree within 0.2\arcsec. However, in each group 
there are other objects with similar right ascension offsets as FIR 4, so it is not 
exceptional. Thus, we conclude that the Spitzer 8, 24 and 70~$\mu$m objects 
are coincident to within the accuracy of our data. 

Similarly, the 160 $\mu$m map was corrected using the offsets derived from 
the 70 $\mu$m observations, given that it was observed at the same time.
There is an offset between the 70 $\mu$m source and the peak of the 160
$\mu$m emission: the brightest part of the extended emission is about 
3\arcsec\ to the northwest of the mid-IR (and 70 $\mu$m) position of 
FIR 4. This 160 $\mu$m peak also overlaps in position with the peak seen 
at 350 and 870 $\mu$m, suggesting that, as opposed to the ``warm'' 
($\leq$ 70 $\mu$m) peak resulting from envelope emission, the ``cold'' 
peak could arise from externally heated dust (see section \ref{clump}).
Although there is a clear peak in the $\lambda$  $\geq$ 160~$\mu$m 
emission towards FIR4, the source in these bands is markedly extended 
and is not a point source. 

We carried out aperture photometry at 70 and 160 $\mu$m by centering
on the peak of the emission in each band.
With aperture photometry at 70 $\mu$m using an aperture radius of 9.6\arcsec, 
a sky annulus of 9.6\arcsec-19.2\arcsec, and an aperture correction factor 
of 1.364, we derived a flux of 132.5 Jy (the sky emission, i.e., the mode of 
fluxes inside the sky annulus, amounted to just 0.1 Jy). 
However, when applying PSF photometry\footnote{We used {\it StarFinder} 
for PSF photometry \citep{diolaiti00}. {\it StarFinder} often finds more than 
one source within 10\arcsec\ of the expected source position. FIR 4 was 
observed four times as part of the HOPS program (4 different groups); in 
two observations {\it StarFinder} found multiple sources. In one of these 
cases, one source was less than 1\arcsec\ from the expected position, 
while the other was 8\arcsec\ away. In the other case {\it StarFinder} 
decomposed the object into 3 nearby sources. To derive the PSF flux, 
we only used the data from three observations: we averaged the fluxes
from the two observations where only one source was found and from the
observation where at least one source was found very close to the expected 
position.}, we derived 40.8 $\pm$ 4.7 Jy (see Ali et al. 2014, in preparation, 
for details). 
It is clear that aperture photometry includes a large contribution of the 
extended, filamentary emission around FIR 4, so PSF photometry should 
yield a more reliable source flux at 70 $\mu$m. The residual images from 
the PSF fits indicated that only a small amount of extended emission was 
left around the position of FIR 4. Thus, the flux value from PSF photometry 
might underestimate the true emission from the protostar at 70 $\mu$m, 
but not by much.

The PACS 160 $\mu$m flux is similarly affected by extended emission.
FIR 4 is brighter than its northern neighbor FIR 3, but it is embedded in the
filament connecting both. Aperture photometry at 160 $\mu$m, centered
on the brightness peak at 160 $\mu$m, with an aperture radius of 12.8\arcsec, 
sky annulus of 12.8\arcsec-25.6\arcsec, and an aperture correction factor 
of 1.515, yielded a flux of 611.2 Jy (once again, the sky emission was low, 
just 1.3 Jy). 
On the other hand, PSF photometry\footnote{Again, {\it StarFinder} 
found several sources in each of the observations of FIR 4; given the 
extended nature of FIR 4 at 160 $\mu$m, we adopted the source flux 
of the brightest object within 8\arcsec\ of the expected position in each
observation and averaged these fluxes.} resulted in 270.2 $\pm$ 8.4 Jy. 
This flux is a slightly better estimate of the envelope emission at 160 $\mu$m,
but given that there is no distinct point source at this wavelength, it probably
still has a large contribution from extended emission. Also, the positional
offset between the 70 and 160 $\mu$m peak indicated that it is possible 
that the main contribution to the 160 $\mu$m emission stems from 
externally heated dust in a massive core that encompasses the protostar 
(see section \ref{clump}), and thus the emission from the envelope itself 
could be very small.
For the PACS 100 $\mu$m flux, we only had aperture photometry available;
adopting the same aperture radius and sky annulus as for the 70 $\mu$m 
data, and an aperture correction factor of 1.440, we derived 287.7 Jy for 
the flux at 100 $\mu$m. Since this value very likely overestimates the 
intrinsic flux from FIR 4, we treat it as an upper limit.

The morphology of OMC-2 FIR 4 is similar at 160 and 350 $\mu$m
(Figure \ref{HOPS108_images}). It is the brightest object in the area, 
but embedded in extended emission. To derive a SABOCA 350 $\mu$m flux 
for this object, we adopted its beam flux of 43.2 Jy (the SABOCA beam has 
a FWHM of 7.34\arcsec), with the centroid determined within a box of size
1.65 $\times$ FWHM around the {\it Spitzer} position to account for potential
offsets. Thus, the beam flux was centered at the brightness peak of the 350 
$\mu$m emission from FIR 4. Aperture photometry with a radius of 3.67\arcsec\ 
(and no sky subtraction) yielded 29.3 Jy; aperture photometry adopting a radius of 
7.34\arcsec\ and a sky annulus of 11.0\arcsec-14.7\arcsec\ resulted in 54.7 Jy.   

At 870 $\mu$m, FIR 4 is clearly the brightest object, but it is also extended
(Figure \ref{HOPS108_images}). The LABOCA 870 $\mu$m beam flux
(beam FWHM of 19\arcsec) amounted to 8.4 Jy; also at 870 $\mu$m, the
centroid position of the source was determined separately, with the {\it
Spitzer} position as starting point. Aperture photometry using a radius half 
the FWHM and no sky subtraction yielded 5.1 Jy, while aperture photometry 
within a radius of 19\arcsec\ and sky annulus of 28.5\arcsec-38\arcsec\ 
resulted in 12.3 Jy. 
\citet{adams12} reported similar SABOCA and LABOCA beam fluxes for FIR 4.

\section{Spectral Energy Distribution and Bolometric Luminosity}
\label{SED}

\begin{figure*}[!t]
\centering
\includegraphics[scale=0.65]{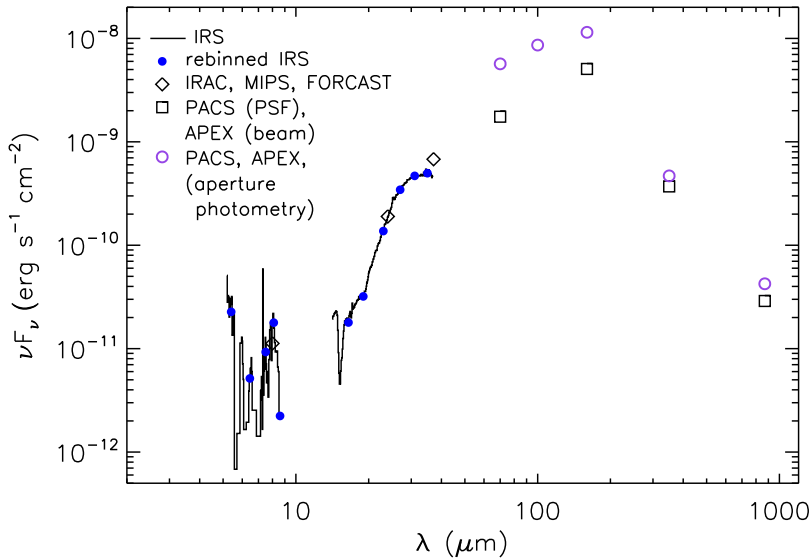}
\caption{SED of OMC-2 FIR 4 using {\it Spitzer} IRAC, IRS, and MIPS,
SOFIA/FORCAST, {\it Herschel} PACS, and APEX SABOCA and LABOCA data. 
The small blue dots on the IRS spectrum show the data points that are used 
in the calculation of $L_{bol}$ and $T_{bol}$ and for the model fits. 
The black diamonds at 8, 24, and 37 \micron\ represent the IRAC, MIPS, 
and FORCAST photometry, respectively. The black squares at 70 and 160 
\micron\ represent PACS PSF photometry measurements, while the open, 
purple circles at these wavelengths and at 100 $\mu$m represent values 
from aperture photometry. The black squares at 350 and 870 \micron\ 
represent SABOCA and LABOCA beam fluxes, while the open, purple circles 
at these wavelengths represent aperture photometry values using radii equal 
to the FWHM of the beams and sky subtraction (see text for details).
\label{HOPS108_SED}}
\end{figure*}

Figure \ref{HOPS108_SED} shows the SED of OMC-2 FIR 4 constructed with
the data mentioned in the previous section. As discussed earlier, the IRS spectrum 
displays deep ice absorption features in the 5-8 $\mu$m region (due to water ice,
methanol, and other organic species), and its silicate absorption feature centered
at 10 $\mu$m is also very deep (with no detected flux over the wavelength
interval of maximum absorption). The CO$_2$ ice feature at 15.2 $\mu$m 
is, in comparison, less pronounced. The change in slope in the spectrum beyond 
about 30 $\mu$m is very likely due to a strong water ice absorption, which is 
fairly broad and centered at $\sim$~45~$\mu$m.
The SED plot also shows the fluxes derived from aperture photometry for
PACS, SABOCA, and LABOCA data. The difference between aperture and
PSF photometry is quite large for PACS fluxes (factors of 3.2 and 2.3 at
70 and 160 $\mu$m, respectively), but smaller at 350 and 870 $\mu$m
(factors of 1.3 and 1.5, respectively). 

As mentioned in section \ref{data}, there seems to be a $\sim$
3\arcsec\ offset in the position of the emission peak between data 
at 8-70 $\mu$m and $\gtrsim$ 160 $\mu$m. When deriving
fluxes using aperture photometry, we re-centered at the peak 
position in each wave band (this is standard procedure for all our
targets in the HOPS sample); thus, at longer wavelengths, the 
aperture was not centered at the same position as was used for 
the fluxes below 100 $\mu$m. If the far-IR and sub-mm emission 
is dominated by an externally heated clump of molecular material 
(which is likely the case; see section \ref{clump}), then our 100, 
160, 350, and 870 $\mu$m fluxes overestimate the emission from 
the protostar itself and should rather be taken as upper limits. 
However, using these fluxes gives an idea of the maximum luminosity 
that could possibly be associated with the protostar OMC-2 FIR 4. 

To calculate the bolometric luminosity ($L_{bol}$) from the observed
SED, we first rebinned the IRS spectrum to exclude those parts of the 
spectrum dominated by ices and to smooth over noisy regions (see Figure 
\ref{HOPS108_SED}). This resulted in flux values at 5.4, 6.45, 7.5, 8.1,
8.6, 16.5, 19.0, 23.0, 27.0, 31.0, and 35.0 $\mu$m that trace the
continuum emission and part of the short-wavelength wing of the broad
silicate absorption feature centered around 10 $\mu$m.
If we use the measured {\it Spitzer} IRAC and MIPS photometry, the
rebinned version of the IRS spectrum, PSF photometry at 70 and 160 $\mu$m,
and the beam fluxes at 350 and 870 $\mu$m, we derive a bolometric luminosity 
of just 36.6 \Lsun. The corresponding bolometric temperature 
($T_{bol}$) is 34.1 K. Neither interpolation of fluxes between the sampled 
values nor extrapolation of fluxes at wavelengths below 5.4 $\mu$m and 
beyond 870 $\mu$m was done. However, even when we extrapolated the
long-wavelength fluxes out to 10 mm using a power law $F_{\nu} 
\propto {\nu}^2$, we derived the same $L_{bol}$ value and
a $T_{bol}$ value that is nearly identical, 34.0 K. Since the fluxes in
the near-infrared are very small, likely much less than 10 mJy, they would
not affect the resulting $L_{bol}$ value.

\begin{deluxetable*}{cllllllll}
\tabletypesize{\footnotesize}  
\tablewidth{0pt}
\tablecaption{{Bolometric Luminosity of OMC-2 FIR 4}
\label{Lbol_table}}
\tablehead{
 & \multicolumn{7}{c}{Data used for $L_{bol}$ calculation}} 
\startdata
{$L_{bol}$ [\Lsun]} & {IRAC} & {MIPS} & {IRS} & {70 $\mu$m} &
{100 $\mu$m} & {160 $\mu$m} & {350 $\mu$m} & {870 $\mu$m} \\
\hline
{\bf 36.6} & {\bf aper} & {\bf PSF} & {\bf yes} & {\bf PSF} & {\bf no} & 
{\bf PSF} & {\bf beam} & {\bf beam} \\
40.7 & aper & PSF & no & PSF & no & PSF & beam & beam \\
78.8 & aper & PSF & yes & aper & aper & aper & beam & beam \\
86.0 & aper & PSF & yes & aper & aper & aper & aper & aper \\
100.2 & aper & PSF & no & aper & aper & aper & aper & aper 
\enddata
\tablecomments{``PSF'' means that the flux was determined via PSF photometry, 
``aper'' means that the flux was measured with aperture photometry, including
subtraction of a sky value determined in a sky annulus. ``no'' means that these 
data were not included in the $L_{bol}$ calculation. Our preferred $L_{bol}$ 
determination is shown in bold.}
\end{deluxetable*}

If we exclude the IRS spectrum from the calculation, we get 40.7 \Lsun\ 
for $L_{bol}$ and 36.5 K for $T_{bol}$. In this case $L_{bol}$ is slightly 
higher, since the interpolated area under the SED is somewhat larger without the 
IRS spectrum.
If we use the mid-IR photometry, IRS spectrum, sub-mm beam fluxes, but
adopt aperture photometry at PACS wavelengths (including the 100 
$\mu$m data point), we calculate 78.8 \Lsun\ for $L_{bol}$ and 36.6 K 
for $T_{bol}$. Finally, adopting aperture photometry at both PACS and APEX 
wavelengths (with the latter using the aperture radius equal to the FWHM of 
the beam and sky subtraction), we derive $L_{bol}$=86.0 \Lsun\ and 
$T_{bol}$=33.8 K (these values change to 100.2 \Lsun\ and 36.7 K, 
respectively, if the IRS spectrum is excluded).
Our $L_{bol}$ calculations are summarized in Table \ref{Lbol_table}.

Thus, depending on which measurements are adopted, we derive bolometric
luminosities ranging from 37 \Lsun\ to 100 \Lsun\ for OMC-2 FIR~4. 
This large range is simply a result of the complex region around this protostar.
However, given that emission from the cold, externally heated clump appears
to dominate the fluxes at $\lambda >$ 160 $\mu$m (see section 
\ref{clump}), the $L_{bol}$ value most closely characterizing the 
protostar is 37 \Lsun.
In contrast, the $T_{bol}$ value shows little dependence on the chosen
photometry, ranging from 34 to 37 K.

\section{Fits of the SED with Standard Protostar Models}
\label{mod}

\begin{figure*}[!]
\centering
\includegraphics[scale=0.72]{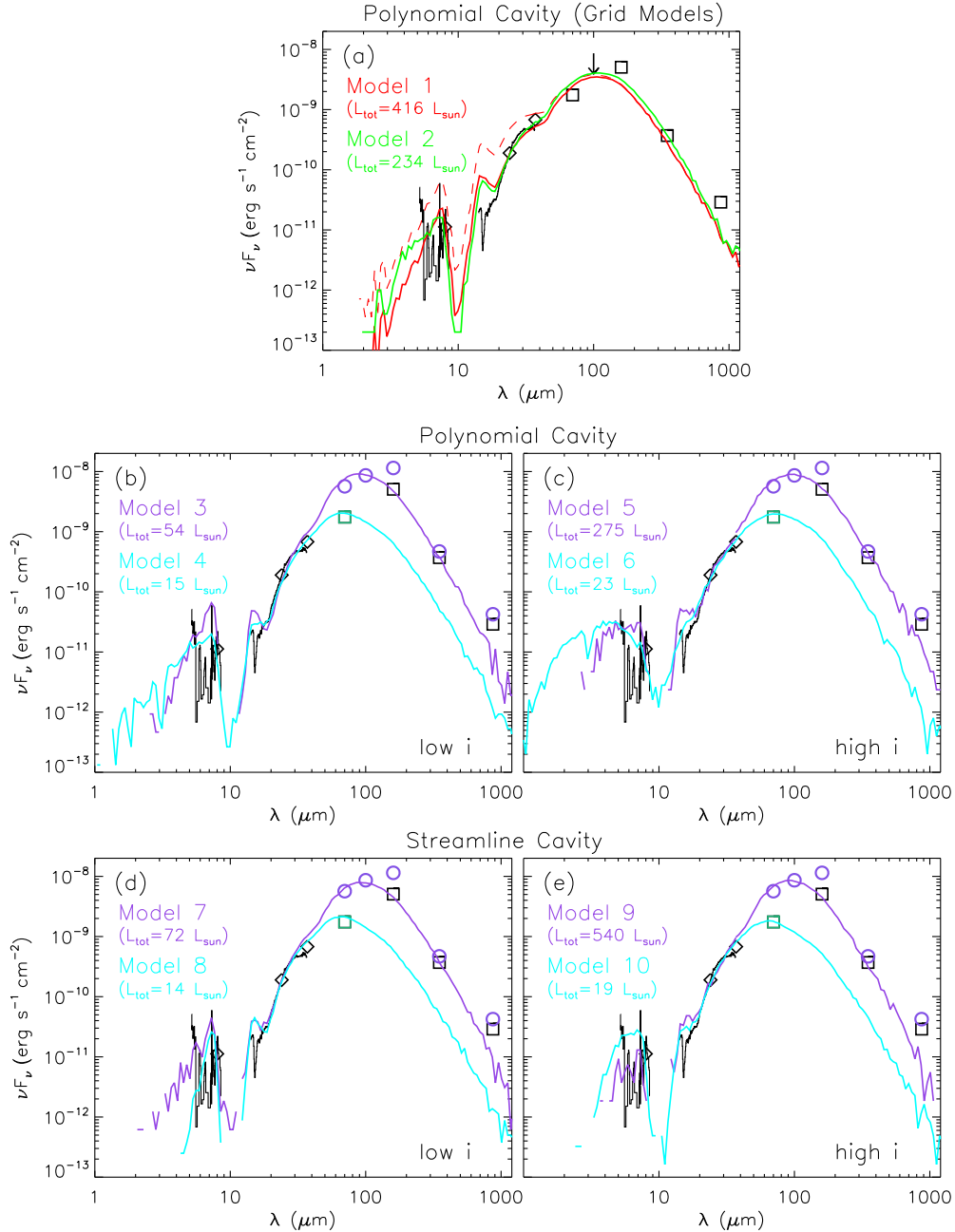}
\caption{SED of OMC-2 FIR 4 from Figure \ref{HOPS108_SED},
shown with ten different model fits (the lines representing the models 
use fluxes from three apertures: 4\arcsec\ for $\lambda < 8\,
\mu$m, 5\arcsec\ for $\lambda =$ 8-37 $\mu$m, and 10\arcsec\ 
for $\lambda > 37\, \mu$m). (a) Best-fit model from the model grid
from Furlan et al. (2014, in preparation), which uses polynomial-shaped
cavities (Model 1, {\it red solid line}; the dashed line shows this same 
model, but with $A_V$ set to 0); best-fit model from the model grid with
$A_V$=0 (Model 2, {\it green line}). 
(b) and (c) Model fits using polynomial-shaped cavities, one set for low
and one set for high inclination angles: fits to the far-IR and sub-mm 
aperture photometry values, shown with the open circles (Models 3 and 
5, {\it purple lines}); fits focusing only on the mid-infrared data and 
the PACS 70 $\mu$m PSF photometry value, shown as the green 
square (Models 4 and 6, {\it cyan lines}). 
(d) and (e) Similar to (b) and (c), but using models with streamline-shaped 
cavities.
See text for details and Table \ref{models} for model parameters.
\label{HOPS108_SED_models}}
\end{figure*}

\citet{adams12} modeled the SED of OMC-2 FIR 4 using IRAC 8.0 $\mu$m,
MIPS 24 $\mu$m, SOFIA/FORCAST 37.1 $\mu$m, PACS 70 and 160 $\mu$m,
and APEX 350 and 870 $\mu$m data. Their reported PACS and APEX fluxes were 
extracted from an earlier version of the reduced maps.
They adopted the sheet collapse model for the envelope from \citet{hartmann96} 
and included the accretion disk model from \citet{dalessio99,dalessio06}. 
The outflow cavities in the envelope were assumed to follow the streamlines of 
infalling particles. Depending on whether they included the 160 $\mu$m data 
point as upper limit, they obtained a total intrinsic luminosity\footnote{The total 
luminosity is the intrinsic luminosity derived from the model fits to the SED; it 
usually differs from the bolometric luminosity, which is derived by integrating
the observed SED and thus depends on, e.g., the inclination angle of the 
source along the line of sight \citep[see, e.g.,][]{whitney03a}.} of 50 \Lsun\ 
(30 \Lsun\ with an upper limit at 160 $\mu$m), an inclination angle of 50\degr\ 
(70\degr), a cavity opening angle of 8\degr, an envelope radius of 5000 AU, 
an envelope reference density $\rho_1$ \citep[see][]{kenyon93}\footnote{The 
reference density $\rho_1$ acts as a scaling factor for the envelope density, 
which determines the thermal emission from the envelope; a $\rho_1$ value 
of $\sim$ 4 $\times 10^{-14}$ g cm$^{-3}$ roughly divides protostars into 
Class 0 and I objects \citep{furlan08, stutz13}.} of 20 (5) $\times 
10^{-13}$ g cm$^{-3}$, and an envelope mass of 10 (2.5) \Msun. 

In Furlan et al. (2014, in preparation), we used models from a grid developed for 
the HOPS program to find the best-fit model of OMC-2 FIR 4 based on an $R$
statistic \citep{fischer12}. These models were calculated using the Monte Carlo 
radiative transfer code developed by \citet{whitney03a,whitney03b}. They use 
the solution of a rotating, infalling cloud core from \citet{terebey84}; the disk 
embedded in the envelope is described by a density power law and a flaring 
angle exponent. The dust opacities were adopted from \citet{ormel11}, 
which include larger, icy grains. As opposed to \citet{adams12}, the cavity 
carved out in the envelope by the outflows was assumed to have a power 
law shape (exponent of 1.5; \citealt{whitney03a}), not conical as in the 
case of a streamline cavity.  
 
\begin{deluxetable*}{lccccccccccc}
\tabletypesize{\footnotesize}  
\tablewidth{0pt}
\tablecaption{{Models for OMC-2 FIR 4}
\label{models}}
\tablehead{
 & &  &  & \colhead{Parameters}  & & & & & & &\\  
\colhead{Model} & \colhead{$L_{tot}$} & 
\colhead{$R_{disk}$} & \colhead{$\rho_1$} & 
\colhead{$\rho_{1000}$} & \colhead{cavity} & 
\colhead{$\theta$} & \colhead{$i$} & \colhead{$A_V$} & 
\colhead{$R$} & \colhead{$L_{bol}$} & \colhead{$T_{bol}$} \\
 & \colhead{[\Lsun]}  & \colhead{[AU]} & \colhead{[g cm$^{-3}$]} &
\colhead{[g cm$^{-3}$]} & & \colhead{[\degr] } & \colhead{[\degr]} & 
\colhead{[mag]} &   & \colhead{[\Lsun]} & \colhead{[K]} 
}
\startdata
Model 1 & 416 & 100 & 7.5 $\times$ 10$^{-13}$ & 2.4 $\times$ 10$^{-17}$  &
 poly & 45 & 70 & 23.9 & 3.81 & 25 & 43 \\
Model 2 & 234 & 500 & 1.9 $\times$ 10$^{-12}$ & 5.9 $\times$ 10$^{-17}$ &
 poly & 45 & 63 & 0.0 & 4.03 & 30 & 41 \\
{\it Model 3} & {\it 54} & {\it 500} & {\it 5.6 $\times$ 10$^{-13}$} & 
{\it 1.8 $\times$ 10$^{-17}$} & {\it poly} & {\it 5} & {\it 32} & {\it 0.0} & 
{\it 6.02} & {\it 58} & {\it 43} \\
Model 4 & 15 & 10 & 1.1 $\times$ 10$^{-13}$ & 3.3 $\times$ 10$^{-18}$ &
 poly & 5 & 49 & 0.0 & 3.03 & 14 & 57 \\
{\it Model 5} & {\it 275} & {\it 500} & {\it 8.3 $\times$ 10$^{-13}$} & 
{\it 2.6 $\times$ 10$^{-17}$} & {\it poly} & {\it 35} & {\it 63} & {\it 0.0} & 
{\it 4.01} & {\it 59} & {\it 42} \\
Model 6 & 23 & 500 & 7.5 $\times$ 10$^{-14}$ & 2.4 $\times$ 10$^{-18}$  &
  poly & 10 & 70 & 0.0 & 3.04 & 14 & 56 \\
{\it Model 7} & {\it 72} & {\it 500} & {\it 7.5 $\times$ 10$^{-13}$} &
{\it 2.4 $\times$ 10$^{-17}$} & {\it stream} & {\it 25} & {\it 32} & {\it 0.0} & 
{\it 4.32} & {\it 52} & {\it 43} \\
Model 8 & 14 & 10 & 9.8 $\times$ 10$^{-14}$ & 3.1 $\times$ 10$^{-18}$  &
 stream & 5 & 49 & 0.0 & 4.41 & 14 & 59 \\
{\it Model 9} & {\it 540} & {\it 500} & {\it 6.0 $\times$ 10$^{-13}$} & 
{\it 1.9 $\times$ 10$^{-17}$} & {\it stream} & {\it 50} & {\it 63} & {\it 0.0} & 
{\it 8.34} & {\it 53} & {\it 43} \\
Model 10 & 19 & 50 & 7.5 $\times$ 10$^{-14}$ & 2.4 $\times$ 10$^{-18}$ & 
 stream  & 15 & 87 & 0.0 & 2.85 & 12 & 60
\enddata
\tablecomments{The model parameters are as follows: $L_{tot}$ is the 
total luminosity (which is the sum of the stellar and accretion luminosity), 
$R_{disk}$ the disk radius (which is equal to the centrifugal radius), 
$\rho_1$ and $\rho_{1000}$ are the reference density at 1 and 1000 AU, 
respectively, $\theta$ is the cavity opening angle, $i$ the inclination angle, 
$A_V$ the foreground extinction along the line of sight, and $R$ is a measure
for the goodness-of-fit. The column labeled ``cavity'' describes the cavity shape: 
``poly'' for polynomial, ``stream'' for streamline. For a polynomial-shaped cavity, 
the cavity shape exponent is 1.5. The $L_{bol}$ and $T_{bol}$ values were 
measured by using the fluxes of the individual model SEDs.
Note that for all models the stellar radius is 6.61 \Rsun, the stellar 
luminosity 10.0 \Lsun, the stellar mass 0.5 \Msun, the disk mass 0.05 \Msun, 
the disk scale height exponent 1.25, and the disk density exponent 2.25. The outer 
envelope radius is set to 10,000 AU. Models in italics are those that, at longer
wavelengths, fit aperture photometry at 70, 100, 160, 350, and 870 $\mu$m, 
while the other models fit the PSF photometry value at 70 $\mu$m (and, in
the case of Models 1 and 2, also PSF photometry at 160 $\mu$m and the
beam fluxes at 350 and 870 $\mu$m). }
\end{deluxetable*}

For the SED fit in Furlan et al. (2014, in preparation), we used the IRAC, MIPS,
rebinned IRS fluxes (as described in section \ref{SED}), the fluxes from
PSF photometry at 70 and 160 $\mu$m, and the beam fluxes at 350 and 870
$\mu$m.
The best-fit model from the grid resulted in a total intrinsic luminosity $L_{tot}$ of 
416 \Lsun, an inclination angle of 70\degr, a cavity opening angle of 45\degr, 
and an envelope reference density $\rho_1$ of 7.5 $\times 10^{-13}$ g 
cm$^{-3}$ (see Model 1 in Figure \ref{HOPS108_SED_models} (a) and Table
\ref{models}). This model also required a substantial foreground extinction of 
$A_V=23.9$. However, the next-best model from the grid resulted in no foreground
extinction, a somewhat lower inclination angle ($i=$63\degr), $L_{tot}$ of 234 
\Lsun, the same cavity opening angle, and a $\rho_1$ value of 1.9 $\times 
10^{-12}$ g cm$^{-3}$ (see Model 2 in Figure \ref{HOPS108_SED_models} 
(a) and Table \ref{models}). Thus, the total luminosity of the source derived 
from models strongly depends on the amount of assumed foreground extinction;
with an $A_V$ value of 23.9 as opposed to 0, the total luminosity changed by
almost a factor of two.

We also ran new model calculations using the same Monte Carlo code of 
\citet{whitney03a,whitney03b} for a more in-depth exploration of the parameter 
space. As a starting point for our input files with the model parameters, we 
used the files from the HOPS model grid; thus, many parameters that we did 
not adjust, such as the stellar mass and disk density exponent, are the same.
We first used a polynomial-shaped cavity, as in our model grid, then a 
streamline-shaped cavity, since the cavity shape can have a large effect 
on the resulting model SED (especially if the cavity opening angle is large), 
but for FIR 4 is not constrained by observations. Similarly, the inclination angle 
is not constrained, so we explored two sets of models, one with lower inclination 
angles and one with a more edge-on orientation.
We also assumed no foreground extinction; substantial extinction along the
line of sight would result in higher total luminosities and change other
model parameters, in particular the inclination angle, given that extinction 
affects mostly the near- and mid-infrared fluxes (see Figure 
\ref{HOPS108_SED_models} (a)).  

In panels (b) and (c) of Figure \ref{HOPS108_SED_models}, we show 
the model fits assuming a polynomial-shaped cavity for more face-on and 
more inclined models, respectively. Each panel shows two model fits each, 
one that considers the long-wavelength aperture photometry (purple lines), 
and one that only tries to fit the mid-infrared data and the PSF photometry 
value at 70 $\mu$m (cyan lines).
As can be seen from Table \ref{models} (Models 3-6) the total luminosity
varies widely, depending on which data points are modeled and which
orientation along the line of sight is assumed. Models that take aperture
photometry at 70 $\mu$m and beyond into account (Models 3 and 5) 
have higher $L_{tot}$ values and higher envelope densities than models that 
consider only the 70 $\mu$m PSF photometry value at long wavelengths 
(Models 4 and 6). Also, more edge-on models require total luminosities that 
are larger than those for models with $i$ $\sim$ 30\degr-50\degr. 
These high-inclination models also have larger cavity opening angles. The 
reference density for models fitting the same data sets changes by about 
50\% between the two sets of inclination angles.

\begin{figure*}[!t]
\centering
\includegraphics[scale=0.74]{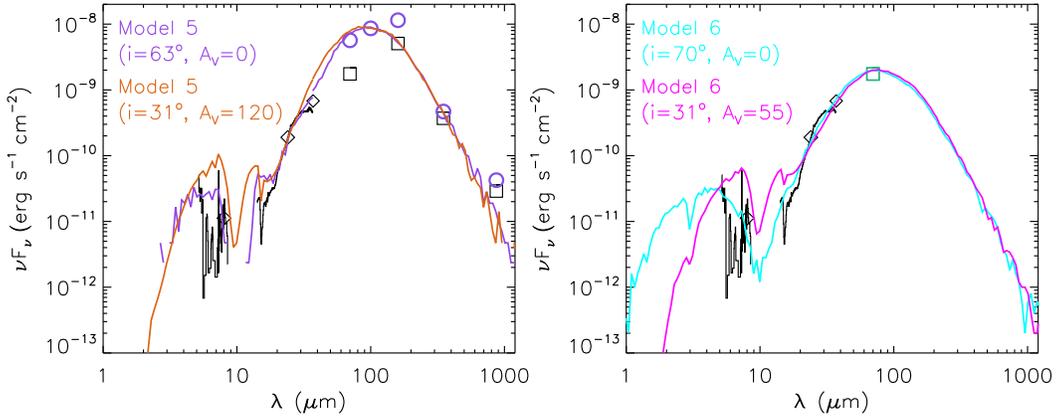}
\caption{SED of OMC-2 FIR 4 with model 5 (left panel) and model 6
(right panel). In each panel, the same model is first shown with the
best-fit inclination angle and $A_V$=0 ({\it purple and cyan lines}), 
then with a low inclination angle and a relatively high foreground extinction
({\it orange and magenta lines}).
\label{HOPS108_models_AV}}
\end{figure*}

\begin{figure}[!t]
\centering
\includegraphics[scale=0.52]{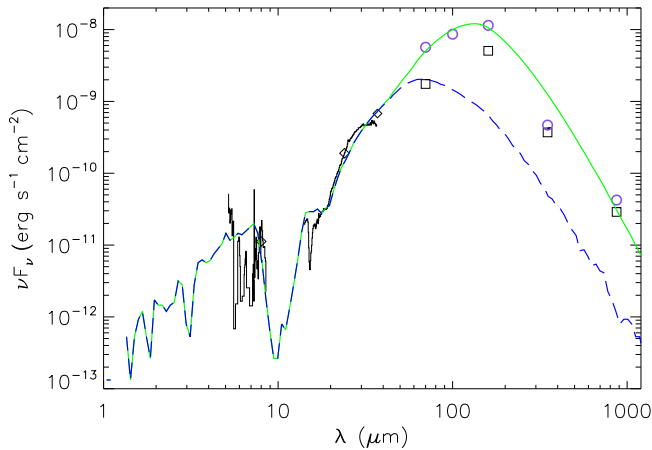}
\caption{SED of OMC-2 FIR 4 as in Figure \ref{HOPS108_SED_models}
(b)-(e) shown with a model composed of Model 4 ({\it blue dashed line};
see also Figure \ref{HOPS108_SED_models} (b)) and a modified blackbody 
with a temperature of 18.5 K; the combined fluxes are shown with the green
solid line.
\label{HOPS108_SED_mBB_model}}
\end{figure}

A similar result applies to the four models that assume a streamline-shaped 
cavity (Figure \ref{HOPS108_SED_models} (d) and (e), Models 7-10
in Table \ref{models}).
Compared to the models with polynomial-shaped cavities, the models
fitting fluxes from aperture photometry at long wavelengths (Models 7 and 9
vs.\ Models 3 and 5) have larger $L_{tot}$ values and cavity opening 
angles, while these parameters are quite similar for the models fitting just
the mid-infrared fluxes and the flux from PSF photometry at 70 $\mu$m 
(Models 8 and 10 vs.\ Models 4 and 6). 
Thus, to reproduce large flux values in the far-IR and a steeply rising SED 
in the mid-IR, a model with a streamline-shaped cavity requires a larger 
cavity opening angle and a much higher total luminosity than models with 
polynomial-shaped cavities. This latter type of cavity evacuates more material 
in the inner envelope than a streamline-shaped cavity, and as a result more 
shorter-wavelength photons can reach the observer. In the extreme case of 
Model 9, where $L_{tot}=$ 540 \Lsun\ is required, the streamline-shaped 
cavity also has to have an opening angle of 50\degr\ such that sufficient 
infrared flux escapes from the envelope.
Interestingly, the models with the streamline-shaped cavity have envelope 
densities similar to those of the models with the polynomial-shaped cavity. 
The higher-luminosity models still require envelopes denser by up to an order
of magnitude.

For each of the models, Table \ref{models} lists their $R$ value, which is
a measure for the goodness-of-fit introduced by \citet{fischer12} (see 
also Furlan et al. 2014, in preparation). $R$ is defined as follows:
\begin{math}
 R = \frac{1}{N} \sum_{i=1}^{N} w_i |\ln(F_{obs}(\lambda_i) -
\ln(F_{mod}(\lambda_i)|,
\end{math}
where $w_i$ is the weight for each data point, $F_{obs}$ and $F_{mod}$
are the observed and model fluxes, respectively, and the sum is over the
number of data points. The weights were set to the inverse of the approximate
fractional uncertainty of each flux measurement and ranged from
1/0.04 to 1/0.4 (with more weight given to the 3-70 $\mu$m region; see
Furlan et al. 2014, in preparation, for details). 
When calculating $R$ values for the various models, we used the same
photometry values and rebinned IRS fluxes (see Figure \ref{HOPS108_SED}
and section \ref{SED}) in the mid-IR, but at longer wavelengths we included 
aperture photometry at $\lambda \geq$ 70 $\mu$m for models 3, 5, 7, 
and 9, and only the 70 $\mu$m PSF photometry at 70 $\mu$m for models 
4, 6, 8, and 10. Thus, the $R$ values for the latter set of models are in 
general lower than those for the former set. Overall, most models have 
$R$ values in the 3-4 range; from a visual inspection of Figure 
\ref{HOPS108_SED_models}, they are indeed quite comparable, 
with the most noticeable differences in the near-IR and in the depth of 
the 10 $\mu$m silicate feature, where there is little to no emission in 
both the observed and modeled fluxes (and thus they do not have a 
measurable effect on the $R$ value).

So far, the models calculated for this work do not include any foreground extinction.
As mentioned earlier and shown in Figure \ref{HOPS108_SED_models} (a), 
extinction will depress the near- and mid-IR fluxes and leave the far-IR
and sub-mm fluxes unchanged. Thus, its effect is similar to that of the
inclination angle; dust in a highly inclined envelope will cause more extinction
of photons on their way to the observer. In Figure \ref{HOPS108_models_AV}
we explored the effect of foreground extinction on two of the high-inclination
models (Models 5 and 6). They are shown with their best-fit inclination
angle (63\degr\ and 70\degr, respectively) and no additional extinction
along the line of sight, and with a low inclination angle of 31\degr\ and 
foreground extinction large enough such that the model fluxes still roughly 
reproduce the observed  SED. With a low inclination angle, model 5 requires 
$A_V$=120, while model 6 needs $A_V$=55. Even though these latter 
models could benefit from tweaking some parameters, they demonstrate 
that inclination angle and foreground extinction are highly degenerate 
parameters. Thus, our models with high luminosity either require a large 
foreground extinction or a more edge-on orientation.

To examine the effect of a significant contribution from extended, externally
heated emission to the far-infrared and submillimeter fluxes, we used a model 
that fits the mid-infrared data points and the PSF photometry at 70 $\mu$m
and added a modified blackbody (for the latter, we used the same dust 
opacities from \citet{ormel11} adopted for our models). All the other models
presented so far do not include such a component; the long-wavelength
fluxes are just fit by emission from the disk and envelope. We chose Model 4 
to represent the protostar, but since Models 4, 6, 8, and 10 all have similar 
SEDs in the 20-1000 $\mu$m range and $L_{bol}$ values of 12-14 \Lsun, 
the choice of model does not affect the results. With the combined protostar 
and blackbody model, we aimed to fit the aperture photometry fluxes at long 
wavelengths, since they are most likely dominated by extended emission. 
The best-fit combination of Model 4 and a modified blackbody is shown in 
Figure \ref{HOPS108_SED_mBB_model}; it requires a blackbody 
temperature of 18.5 K. The PACS 70, 100, and 160 $\mu$m fluxes
are fit well, while the 350 $\mu$m flux is overestimated and the 870
$\mu$m flux slightly underestimated. The discrepancy at 350 $\mu$m
is likely an aperture effect, given the small aperture used at this wavelength.
The combined bolometric luminosity of Model 4 and the 18.5 K blackbody
amounts to 76 \Lsun, which is very similar to the $L_{bol}$ value 
measured from the observed SED with the mid-IR data, aperture photometry 
at 70, 100, and 160 $\mu$m, and the beam fluxes in the sub-mm (see 
section \ref{SED}). The contribution of the modified blackbody to this
$L_{bol}$ value is 62 \Lsun, leaving just 14 \Lsun\ as $L_{bol}$ for
the protostar.

\section{Discussion}

\subsection{The Bolometric Luminosity of the Protostar OMC-2 FIR 4}
\label{Lbol_discussion}

Our new measurements of OMC-2 FIR 4 in the far-IR and submillimeter
suggest that its bolometric luminosity is far below the most recent value
of $\sim$ 1000 \Lsun\ suggested in the literature: we derive a range
from 37 \Lsun\ to $\sim$ 100 \Lsun, with the former value more 
likely to describe the protostar, since it is derived using smaller apertures 
for the source fluxes. Also, given that the IRS spectrum plays an important 
role in constraining the mid-infrared part of the SED, the more realistic upper
limit for $L_{bol}$ is 86 \Lsun\ (which is derived when including the IRS
spectrum in the SED).  

As noted by \citet{lopez13}, the fluxes and thus luminosity of OMC-2
FIR 4 depend strongly on which apertures are used. The larger the
aperture, the more envelope emission, but also extended emission
from the surrounding filament and emission from neighboring sources
is included. 
We find that, for isolated protostars in the Orion star-forming region
(distance of $\sim$ 420 pc; \citealt{menten07, hirota07}), 
aperture radii of $\sim$ 10\arcsec\ in the far-IR (70-160 $\mu$m) 
capture most of the emission from the envelope at these wavelengths 
(see Furlan et al. 2014, in preparation). Choosing larger radii risks 
including surrounding emission that is not associated with the envelope. 
\citet{crimier09} likely overestimated the fluxes of FIR 4, since they 
integrated their derived continuum profiles at 350, 450, and 850
$\mu$m out to $\sim$ 20\arcsec, their derived envelope size. 
However, FIR 4 seems to be surrounded by extended emission, and
a 20\arcsec\ aperture will include emission from this surrounding
material. Their derived fluxes at 350 and 850 $\mu$m are about
an order of magnitude larger than our APEX fluxes at similar wavelengths.
\citet{crimier09} also extracted IRAS fluxes at 60 and
100 $\mu$m; however, the IRAS beam is very large at these
wavelengths, and contamination by FIR 3 and FIR 5 likely also plays
a role. Their extracted MIPS 24 $\mu$m flux is also overestimated due to
the aperture radius of 15\arcsec; their flux value of 5.0 Jy is 3.3 times
larger than the value derived by \citet{megeath12} with PSF fitting.

The big discrepancy in flux measurements resulting from adopting
apertures that are at most a factor of a few different suggests that
the region around FIR 4 is very complex and contains copious amounts
of extended emission. The dust in this extended material is likely to be 
heated by the strong far-IR radiation field present in the Orion cloud 
complex, and is not internally heated by the protostar itself. Thus, if we 
want to characterize the protostar itself, it seems reasonable to adopt 
conservative (i.e., smaller) aperture sizes to measure fluxes. The offset 
we found for the emission peak at $\lambda$ $\leq$ 70 $\mu$m and 
$\lambda$ $\geq$ 160 $\mu$m suggests that even our lowest flux 
values in the 160-870 $\mu$m region could overestimate the envelope 
emission. This is supported by the interferometer results of \citet{lopez13},
who found that there are multiple components in OMC-2 FIR 4, only one 
of which contains a protostar. We therefore ignore the more extended core 
resolved in the observations of \citet{shimajiri08} and \citet{lopez13} and 
focus on the properties of the protostar and its inner ($\lesssim$ 8,000 AU 
radius) envelope. As shown in section \ref{mod}, protostellar models that
assume a dense, infalling envelope around a single protostar can
describe the observed SED. We will discuss the relationship of this
protostar to the more extended core in the FIR 4 region in section 
\ref{clump}.

\subsection{The Classification of the Protostar OMC-2 FIR 4}

Adopting an $L_{bol}$ value of 37 \Lsun\ for FIR 4, and using the 
beam fluxes at 350 and 870 $\mu$m, we can calculate the ratio of
sub-mm luminosity ($L_{submm}$) and $L_{bol}$. For $L_{submm}$,
we integrated the SED at wavelengths $\geq$ 350 $\mu$m 
\citep{andre93}. We derived $L_{submm}$/$L_{bol}$ of 2\%
(the result is the same if we add a long-wavelength extrapolation to
the SED reaching to 10 mm). This value is four times larger than the
minimum value for a Class 0 protostar \citep{andre93}, so OMC-2 FIR 4
appears to be in a very early, evolutionary state, when presumably most 
of the stellar mass is still in the envelope or core surrounding the protostar.
Thus, it shares some of the properties of the PACS Bright Red sources 
(PBRs) discovered by \citet{stutz13}, which are among the youngest 
protostars, with high envelope densities and infall rates. The log of the ratio 
of its 70 and 24 $\mu$m flux in $\lambda F_{\lambda}$ space 
amounts to 0.96, while for PBRs this ratio is larger than 1.65. Nonetheless,
our model fits (section \ref{mod}) yielded high envelope densities,
$\rho_1$ $\geq$ 7.5 $\times$ 10$^{-14}$ g cm$^{-3}$, very 
similar to those of the PBRs studied in \citet{stutz13}. 

Even model fits that at longer wavelengths only included the 70 $\mu$m 
PSF photometry point (Models 4, 6, 8, and 10) resulted in $\rho_1$ values 
close to 1.0 $\times$ 10$^{-13}$ g cm$^{-3}$, suggesting that even if 
we assume that the far-IR and sub-mm emission is dominated by externally
heated dust, the derived envelope density of FIR 4 is still large. These 
models also yielded $L_{bol}$ values of 12-14 \Lsun\ for the protostar, 
which is less than half the smallest value measured from the observed SED, 
but is a result of these model fluxes being lower by almost a factor of 10 
compared to the beam fluxes at 350 and 870 $\mu$m and the flux from PSF 
photometry at 160 $\mu$m (see Figure \ref{HOPS108_SED_models}).
The ratios of sub-mm to bolometric luminosity for these models is
0.5-0.6\%, on the low end for a Class 0 object \citep{andre93}, but
the $T_{bol}$ values are $\leq$ 60 K and thus clearly in the Class 0 
range \citep{chen95}.
Therefore, there is strong observational evidence that OMC-2 FIR 4 is a 
Class 0 protostar. The high envelope density suggests that it is in an
early evolutionary stage, and its SED classification as a Class 0 object
translates into a Stage 0 physical state \citep[see][]{robitaille06}.

\subsection{Determining Source Properties from Models}

As shown in section \ref{mod}, a wide range of models can fit the observed
fluxes of OMC-2 FIR 4, especially due to the somewhat uncertain source 
fluxes in the far-infrared and submillimeter and unconstrained model 
parameters such as the inclination angle and cavity shape. On the other hand, 
the current model fits can already give rough estimates for some source properties:
the envelope density is relatively high, in the $\rho_1$ $\sim$ 10$^{-13}$
$-$ 10$^{-12}$ g cm$^{-3}$ (or $\rho_{1000}$ $\sim$ 3 $\times$ 
10$^{-18}$ $-$ 3 $\times$ 10$^{-17}$ g cm$^{-3}$) range, and 
the cavity is either relatively narrow, combined with a more face-on orientation, 
or wide, if the inclination angle is high or a large amount of foreground extinction
is present.

When comparing our models 7 and 9 to the modeling results of \citet{adams12},
the envelope reference densities are similar, but our total luminosities and
cavity opening angles are larger. Even though all these models assume 
streamline-shaped cavities, \citet{adams12} used the sheet collapse solution 
for the envelope structure. This results in a wider region of decreased density 
along the rotation axis (which is the same as the outflow axis) compared to our 
TSC models, roughly corresponding to a wider cavity. Thus, a sheet-collapse 
model with a smaller cavity still allows the escape of a large amount of mid-IR 
photons, even at higher inclination angles. Our models require both higher total 
luminosities (by up to a factor of $\sim$ 20) and larger cavity opening angles 
(by up to a factor of 6) to allow sufficient photons to reach the observer. 

Inferring the total luminosity for FIR 4 is more difficult. While the bolometric 
luminosity is derived from the observed SED (assuming isotropic emission), 
the total luminosity is the intrinsic energy output from the object. Depending
on the inclination angle, cavity opening angle, or the amount of foreground 
extinction, an object with the same $L_{tot}$ value will have $L_{bol}$ 
values that are higher or lower \citep[see][]{whitney03a}. Observed fluxes 
determine $L_{bol}$, so it is not straightforward to convert it to an L$_{tot}$ value.
With $A_V=23.9$, $L_{tot}$ of FIR 4 can be as high as 416 \Lsun, but 
also a model with a streamline-shaped cavity, $A_V=0$, and a more 
edge-on orientation can yield $L_{tot}=$ 540 \Lsun. On the other hand,
the $L_{bol}$ value derived from these two model SEDs is 25 \Lsun\ for the
former model and 53 \Lsun\ for the latter one (see Table \ref{models}).

\begin{figure}[!t]
\centering
\includegraphics[scale=0.52]{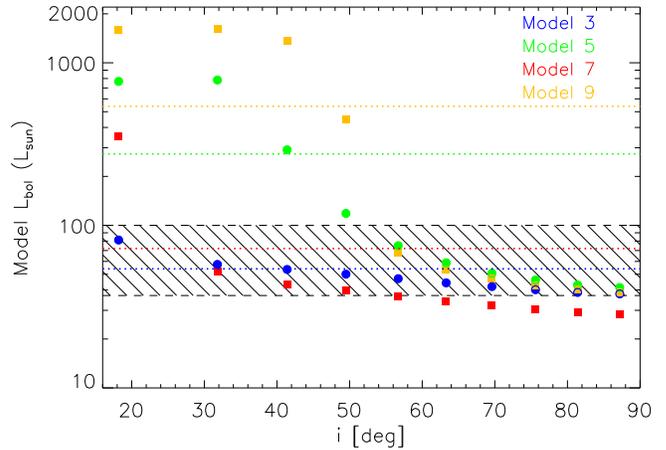}
\caption{The bolometric luminosity versus the inclination angle for Models
3 ({\it blue}; $L_{tot}$=54 \Lsun), 5 ({\it green}; $L_{tot}$=275 \Lsun), 
7 ({\it red}; $L_{tot}$=72 \Lsun), and 9 ({\it yellow}; $L_{tot}$=540 \Lsun).
The horizontal, dotted lines show the value of $L_{tot}$ for these four
models. The dashed region shows the range of $L_{bol}$ values 
derived from the observed SED of OMC-2 FIR 4 (see text for details). 
\label{HOPS108_model_Lbol_inc}}
\end{figure}

\begin{figure}[!t]
\centering
\includegraphics[scale=0.52]{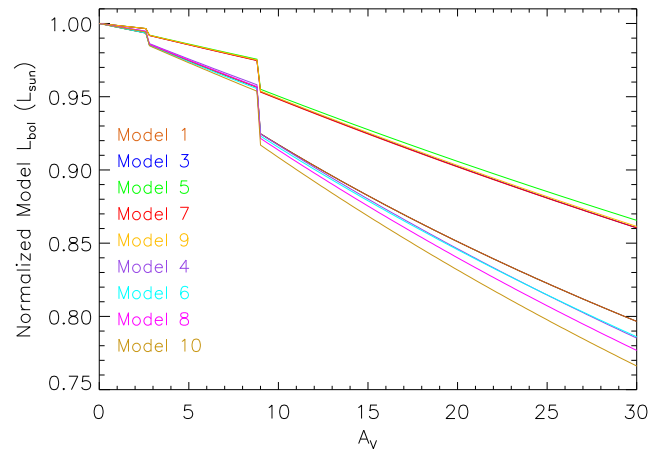}
\caption{The bolometric luminosity, obtained after the fluxes were 
attenuated by extinction and normalized to the $L_{bol}$ value at $A_V$=0, 
versus the foreground extinction $A_V$ for Models 1 ({\it orange}), 
3 ({\it blue}), 5 ({\it green}), 7 ({\it red}), and 9 ({\it yellow}), 
4 ({\it purple}), 6 ({\it cyan}), 8 ({\it magenta}), and 10 ({\it beige}). 
The discontinuities at $A_V$ values of 2.7 and 9 are due to the transitions 
to different extinction laws (\citealt{mathis90} $R_V$=5 curve for 
$A_V$=0-2.7, and two curves from \citealt{mcclure09} for 
$A_V$=2.7-9 and 9-30). 
\label{HOPS108_model_Lbol_AV}}
\end{figure}

In Figure \ref{HOPS108_model_Lbol_inc}, we show the effect of inclination
angle on $L_{bol}$ derived for the fluxes of those models that aim at fitting 
the aperture photometry values at longer wavelengths (Models 3, 5, 7, and 9). 
Each model has a certain total luminosity, $L_{tot}$ (see Table \ref{models}), 
which does not depend on the inclination angle. However, $L_{bol}$, which is 
derived from the SED, strongly depends on the viewing angle \citep[see]
[section 3.3]{whitney03a}. For more face-on orientations, $L_{bol}$ is higher 
than $L_{tot}$, especially for the higher-luminosity models. At $i \gtrsim$ 45\degr, 
the bolometric luminosity is lower than $L_{tot}$. Thus, to fit the observed 
$L_{bol}$ values, a high-luminosity model requires a larger inclination angle 
than a model with a lower total luminosity. This is reflected in the results 
presented in Table \ref{models}, where the models with the highest luminosity 
have inclination angles $\geq$ 63\degr. Given that $L_{bol}$ of FIR 4 is likely 
$\sim$~40~\Lsun\ based on the observations (see section \ref{SED}), a total 
luminosity of a few hundred \Lsun\ is only possible if the inclination angle of 
the object is relatively high.

The effect of foreground extinction on $L_{bol}$ is less dramatic than
that of the inclination angle, as long as $A_V \lesssim$ 30 (see Figure 
\ref{HOPS108_model_Lbol_AV}). 
Using all the models from Table \ref{models} except for Model 2, we calculated 
$L_{bol}$ for $A_V=$ 0-30 (i.e., we extinguished the fluxes using $A_V$
values ranging from 0 to 30, then computed $L_{bol}$) and plotted $L_{bol}$ 
normalized by its value at $A_V$=0. Models 3, 5, 7, and 9 (which aimed at 
fitting the higher far-IR and sub-mm flux values) show a nearly identical 
decline in $L_{bol}$ with increasing extinction; at $A_V$=30, $L_{bol}$ is 
86\% of its value at $A_V$=0. The $L_{bol}$ values for models 4, 6, 8, 
and 10 (which fitted just the 70 $\mu$m PSF photometry at longer 
wavelengths) decrease more steeply, reaching 76-80\% of their
extinction-free values at $A_V$=30.
The best-fit model from the grid (Model 1) is the only model that included
foreground extinction in order to fit the SED; with $A_V$=0, its
$L_{bol}$ value is 29 \Lsun, and for the best-fit $A_V$ of 23.9, 
$L_{bol}$ decreases to 25 \Lsun. As mentioned in section \ref{mod},
extinction and inclination angle are degenerate parameters; some of our
models with $A_V$=0 and a large inclination angle could actually be
modified to models with lower inclination angles and larger $A_V$ values, 
with hardly any change in $L_{bol}$, given that $L_{bol}$ increases 
for lower inclination angles, but decreases with $A_V$. 

The total luminosity also depends on which data sets are fit. While the near- and 
mid-infrared fluxes of the models in Figure \ref{HOPS108_SED_models} 
(b) to (e) are comparable, they are strikingly different in the far-infrared and 
submillimeter. If we adopt the fluxes from aperture photometry to represent 
the emission from the envelope at long wavelengths, the envelope is 
5-11 times denser than if we use the flux from PSF photometry at 
70 $\mu$m. The difference in internal luminosity when fitting these two
data sets is a factor of 4-5 for the low-inclination models, but increases to 
a factor of 12-28 for the high-inclination models. When PSF photometry 
at 70 $\mu$m is used, the model fluxes beyond 100 $\mu$m seem to 
be too low by about an order of magnitude. 
The discrepancy between model and data at 160, 350, and 870 $\mu$m 
could be explained if the PSF photometry at 160 $\mu$m and the 
beam fluxes in the sub-mm were contaminated by extended emission or 
if the PSF photometry at 70 $\mu$m underestimated the true envelope 
emission.

We showed that a protostellar model that only considers fluxes out to 
70 $\mu$m (using the PSF photometry value at that wavelength)
and adds contribution of $\sim$~20~K dust can reproduce the 
SED that uses aperture photometry fluxes at long wavelengths. In
this case, most of the far-IR and sub-mm emission is generated by
this extended dust component that is not necessarily part of the 
envelope when modeling FIR 4. Given that its bolometric luminosity
is 62 \Lsun, compared to 14 \Lsun\ for the protostar, the dust 
must be heated by external sources (see also section
\ref{Lbol_discussion}). Thus, if this interpretation of the SED is correct, 
the protostar associated with OMC-2 FIR 4 is of just moderate luminosity.

\subsection{Does OMC-2 FIR 4 have an outflow?}
\label{outflow}

A key piece of evidence to support the interpretation of OMC-2 FIR 4 as 
a young protostar would be the detection of an outflow. While OMC-2 FIR 3 
drives an outflow that likely reaches FIR 4, it is not clear whether FIR 4 itself
powers one.

\begin{figure}[!t]
\centering
\includegraphics[scale=0.49]{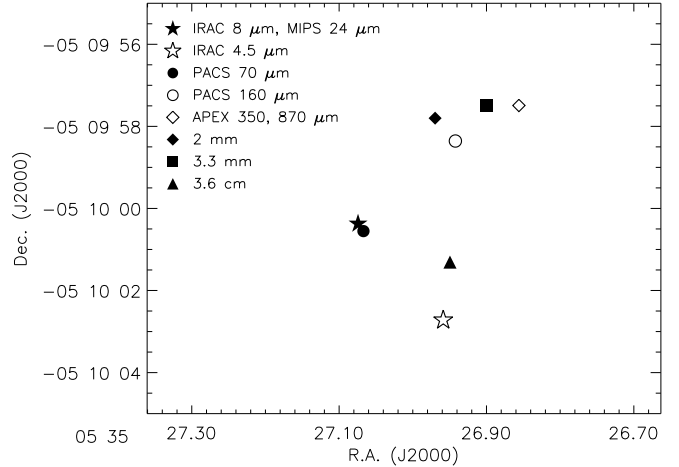}
\caption{The positions of OMC-2 FIR 4 in the sky in different wave
bands. At 4.5 $\mu$m, the position shown is that of the offset peak
seen in Figure \ref{HOPS108_images}. The positions at 70 and 160 
$\mu$m are those of FIR 4 from scan group 130, corrected by the
average offset of HOPS sources in that group relative to their {\it Spitzer}
positions. The 2 mm, 3.3 mm, and 3.6 cm positions are taken from
\citet{lopez13}, \citet{shimajiri08}, and \citet{reipurth99}, respectively.
The coordinate values are listed in Table \ref{FIR4_positions}.
\label{HOPS108_positions}}
\end{figure}

VLA imaging by \citet{reipurth99} detected an elongated radio continuum source
toward OMC-2 FIR 4 at 3.6 cm; it is the weakest of the three centimeter sources 
detected in its vicinity, the others are FIR 3 (SOF 2N, HOPS 370) and SOF 5 
(also HOPS 368; \citealt{adams12}). They interpreted the centimeter 
source as free-free emission from shocks in an outflow driven by FIR 4; this is the 
favored interpretation for centimeter radio jets found towards low mass protostars 
\citep{anglada96}.
The location of the radio continuum source is consistent with the strong emission 
peak at 4.5 $\mu$m and a weaker one at 5.8 $\mu$m, which appear in
our data offset relative to the peak position seen at 8-70 $\mu$m (see
Figure \ref{HOPS108_positions} and Table \ref{FIR4_positions}). Outflow knots 
can be detected at 4.5 $\mu$m and 5.8 $\mu$m due to shocked $H_2$ emission 
\citep{smith05}, so it is possible that the southwestern lobe of an outflow from 
FIR 4 has been detected also in the mid-IR. Alternatively, the 4.5 and 5.8 
$\mu$m emission could be light scattered in an outflow cavity \citep{whitney03a}. 

\begin{deluxetable*}{cccc}
\tabletypesize{\footnotesize}  
\tablewidth{0pt}
\tablecaption{{Positions of the Peak Emission from OMC-2 FIR 4 at Different 
Wavelengths}
\label{FIR4_positions}}
\tablehead{
\colhead{Wavelength} & \colhead{R.A. (J2000)} & 
\colhead{Dec. (J2000)} & \colhead{Reference}}
\startdata
4.5 $\mu$m & 5  35  26.96 & -5  10  02.7 & \citet{megeath12} \\
8.0, 24 $\mu$m & 5  35  27.07 & -5  10  00.4 & \citet{megeath12} \\
70 $\mu$m & 5  35  27.07 & -5  10  00.6 & this work \\
160 $\mu$m & 5  35  26.94 & -5  09  58.4 & this work \\
350, 870 $\mu$m & 5  35  26.85 & -5  09  57.5 & this work \\
2.0 mm & 5  35  26.97 &  -5  09  57.8 & \citet{lopez13} \\
3.3 mm &  5  35  26.90 &   -5  09  57.5 & \citet{shimajiri08} \\
3.6 cm & 5  35  26.95  &  -5  10  01.3 & \citet{reipurth99}
\enddata
\tablecomments{These positions are shown in Figure 
\ref{HOPS108_positions}; see the figure caption for notes on
the 4.5, 70, and 160 $\mu$m positions.}
\end{deluxetable*}

\citet{anglada95,anglada96} showed that the centimeter flux from 
low- to intermediate-mass young stars is dominated by collisional ionization 
in outflow-driven shocks. They found that the centimeter luminosity is 
correlated with the momentum rate in the outflow ($\dot{P}$), which 
is in turn correlated with $L_{bol}$. The resulting empirical relationship 
between the centimeter luminosity and $L_{bol}$ for low- to intermediate-mass
stars is $S_{\nu} d^2/(\mathrm{mJy\: kpc}^2) = 10^{-2.1} 
(L_{bol}/L_{\odot})^{0.6}$, where $S_{\nu}$ is the cm flux in mJy 
and $d$ is the distance to the source in kpc (G. Anglada, private 
communication). 
Using the flux densities in \citet{reipurth99} and adopting a distance of 420 pc, 
we find $L_{bol}$ values of $\sim$ 800, 80, and 190 \Lsun\ for FIR 3, FIR 4, 
and SOF 5, respectively. Thus, the radio continuum flux is most consistent with 
a $\sim 100$~\Lsun\ luminosity for FIR 4, although there is considerable scatter 
and uncertainty in the relationship relating luminosity to centimeter flux. Just based
on its $S_{\nu} d^2$ value of 0.11, the centimeter emission from FIR 4 is consistent
with either a jet from a $\sim$ 100 \Lsun\ source (shock ionization) or an 
HII region from a $\gtrsim$ 10$^3$ \Lsun\ source (photoionization) (see 
Figure 5 of \citealt{anglada95}). Our data and models are in favor of the
lower luminosity for FIR 4, thus supporting the interpretation of the centimeter
emission as originating in an outflow. Given that $\dot{P} \propto (S_{\nu} 
d^2)^{1.1}$ \citep{anglada95}, we can also estimate that the outflow of FIR 4 
has four times less momentum flux than the outflow driven by FIR 3 
(0.64 mJy for FIR 4 vs.\ 2.48 mJy for FIR 3 at 3.6 cm). 
 
Alternatively, \citet{lopez13} interpreted the centimeter emission as arising from
photoionization from an embedded B3-B4 ZAMS star, which would have a luminosity 
of 600 \Lsun\ to 1000 \Lsun. We do not favor this interpretation since there is no
supporting evidence for photoionization; the IRS spectrum of the source does not detect
the polycyclic aromatic hydrocarbon (PAH) features that are common in reflection 
nebulae around intermediate-luminosity stars.  However, given the high extinction 
towards the source and the broad range of plausible luminosities for FIR 4, it is 
not currently possible to rule out this interpretation. 

Additional evidence for an outflow comes from the far-IR CO spectra. Recently,
{\it Herschel} spectroscopy with PACS has shown that FIR 4 has the highest 
far-IR CO luminosity in the sample of Orion protostars studied by \citet{manoj13}. 
The high-excitation, far-IR molecular line emission appears compact ($\sim$ 2000 AU) 
and centered on FIR 4. The excitation energies and critical densities of the transitions
suggest that the far-IR CO emission originates in hot (T $>$ 300 K) gas, with
fits to the far-IR CO rotational excitation diagrams yielding temperatures exceeding
2000 K and relatively low densities of n(H$_2$) $<$ $10^6$ cm$^{-3}$
\citep{manoj13}. This hot gas is likely heated by shocks, possibly inside the 
outflow cavity or along cavity walls. 
\citet{kama13} detected broad wings in the far-infrared lines of OH, H$_2$O, 
and CO observed with {\it Herschel}/HIFI. The line wings were symmetric 
and their widths and strength increased with the excitation level of the line,
suggesting emitting gas that is hot or dense and thus possibly a compact 
outflow from FIR 4 that contributes to the CO emission. When combining their 
results with those of \citet{manoj13}, \citet{kama13} derived a total CO 
luminosity of $\sim$~0.4 \Lsun. If we assume an $L_{bol}$ value of 
37 \Lsun, the CO luminosity amounts to 1\% of the total energy output. 
However, it is not clear whether the protostar associated with FIR 4 is responsible
for the entire CO emission. Using interferometric observations, \citet{shimajiri08} 
argued that the high velocity CO emission toward FIR 4 is due to the outflow 
from FIR 3 colliding with the FIR 4 clump. In their interpretation, the morphology 
of the FIR 4 core arises from the interaction with the FIR 3 outflow 
\citep[also see][]{lopez13}. In this case, the far-IR CO lines may have a 
significant contribution from the shock driven by FIR 3.

\subsection{The molecular clump associated with OMC-2 FIR 4}
\label{clump}
  
Besides the offset between the 8-70 $\mu$m peak and the peak at 
4.5 $\mu$m, Figure \ref{HOPS108_positions} also shows the offset 
between the 70 and 160 $\mu$m peaks mentioned in section \ref{data}.
The 160 $\mu$m peak also roughly coincides with the peak position in
our APEX data and those reported at 2 mm \citep{lopez13} and 3.3 mm 
\citep{shimajiri08} (see also Table \ref{FIR4_positions}). Thus, the more 
extended emission at $\lambda$ $\geq$ 160 $\mu$m probably probes 
a dense clump of dust and gas heated externally. This scenario is similar 
to the Bok globule studied by \citet{stutz10} containing a protostar and 
a starless core, with the emission from the latter source starting to be 
noticeable at $\lambda$ $>$ 100 $\mu$m and becoming comparable 
to the emission from the protostar in the sub-mm. 

To examine the properties of this molecular core, we integrated the flux 
in a 20\arcsec\ aperture centered on the core position, using an annulus 
from 30\arcsec\ to 40\arcsec\ to subtract out the more extended 
emission. For the core position, we used the average centroid position 
from the 160, 350 and 870~$\mu$m maps ($\alpha$ (J2000) = 
5$^h$ 35$^m$ 26.88$^s$ and $\delta$ (J2000) = -05\degr\ 
9\arcmin\ 57.8\arcsec). After applying the aperture corrections for 
a point source (which primarily account for flux scattered to large angles 
in the 160~$\mu$m data), we fit a modified blackbody to the 160-870 
$\mu$m photometry using the same opacity law from \citet{ormel11} 
as for the HOPS model grid and our models. The resulting fit gives a 
temperature of 22 K, a mass of 27.3~\Msun, and an overall luminosity 
of 137~\Lsun. 
This is consistent with our earlier model that combined a protostellar
model and a modified blackbody fit to our standard fluxes from aperture
photometry (with aperture radii $<$ 20\arcsec; thus, our earlier fit 
yielded a somewhat smaller temperature of 18.5 K). 

Our analysis shows that the FIR 4 core is massive with a mass commensurate 
with that of a high-mass star.  Although the possibility of higher temperatures 
due to the heating by the protostar may affect the measurement, the localized
mass should be within a factor of two of the observed mass and therefore 
would still be in excess of 10~\Msun.  Our mass is consistent with the masses 
determined from the interferometer data by \citet{lopez13} for a range of 
assumed temperatures. \citet{li13} derived a core mass of 13 \Msun\ based 
on NH$_3$ data; they inferred that the FIR 4 core was just massive enough
to be in virial equilibrium and thus gravitationally bound.
The placement of the protostar near the edge of the sub-mm clump is consistent 
with the claim of \citet{lopez13} and \citet{shimajiri08} that the sub-mm core 
has fragmented and is forming multiple objects, with the observed protostar to be 
the first object to form with a high enough luminosity to be detected.  We also 
note that due to the large reservoir of gas mass associated with the protostar, 
although the observed protostar currently appears to have a modest luminosity, 
it may continue to grow in mass and luminosity as it draws from the core. 
Hence, the final mass of the protostar, and whether it may form a low-mass 
star or intermediate-mass star, is highly uncertain.  

Given that the long-wavelength ($\lambda \geq$ 100 $\mu$m) emission 
from FIR is dominated by the massive core that is mostly heated externally, 
the protostar itself is probably best described by our models that fit the mid-IR 
data and the PSF photometry at 70 $\mu$m and resulted in $L_{tot}$ values 
ranging from 14 to 23 \Lsun\ (see section \ref{mod} and Table \ref{models}). 
Also, given the absence of a wide outflow from FIR 4, SED models that require a 
high total luminosity and large cavity opening angles seem more unrealistic. 
A young, spatially compact outflow would not have had sufficient time to carve 
a large cavity within the envelope. Among the low-luminosity models (Models 
4, 6, 8, and 10), those with a larger inclination angle have wider cavities and 
somewhat larger $L_{tot}$ values (such that sufficient mid-infrared photons
can still reach the observer despite the high inclination), so we favor the two 
models with $i=49$\degr. In addition, the models with lower inclination angles (which 
typically require larger envelope densities) seem to result in deeper silicate 
absorption features and lower near-infrared fluxes, which better matches 
the observations.

\section{Conclusions}

OMC-2 FIR 4 is an intriguing protostar whose nature has been debated
in the literature; it is likely deeply embedded and thus in an early 
evolutionary stage, but its properties, like luminosity and envelope mass, 
were poorly determined. We clearly detect protostellar emission at 
$\lambda$ $\leq$ 70 $\mu$m, but at longer wavelengths the larger 
molecular core dominates the emission. We present the most complete 
analysis to date of this object. Using data from the {\it Spitzer}, 
{\it Herschel}, and APEX telescopes, we derive new values for the bolometric 
luminosity of OMC-2 FIR 4 and estimate some of its envelope properties from 
model fits. Some ambiguities on the detailed nature remain due to the deeply 
embedded state of the protostar.
Our main conclusions are as follows:
\begin{itemize}
\item We construct the SED of OMC-2 FIR 4 with photometry at 8, 24, 
37.1, 70, 100, 160, 350, and 870 $\mu$m, and spectroscopy from 5 to
37 $\mu$m. Thus, the SED is well-sampled, in particular at wavelengths
where the emission peaks. We obtain more accurate photometry of the 
protostar and its envelope by choosing smaller apertures ($\sim$ 
10\arcsec) in the 70-870 $\mu$m range than were previously adopted. 
However, we note an offset of $\sim$~3\arcsec\ in the emission peak 
for $\lambda$ $\leq$ 70 $\mu$m and $\lambda$ $\geq$ 160 
$\mu$m, which suggests that at long wavelengths we actually 
probe a clump of externally heated dust and thus even our fluxes at 
160, 350, and 870 $\mu$m could overestimate the envelope emission.
\item The bolometric luminosity of OMC-2 FIR 4 ranges from 37 \Lsun\
to 100 \Lsun, depending on which values are adopted for the far-IR
and sub-mm photometry. Given that the extended emission surrounding 
this object at long wavelengths ($\gtrsim$ 70 $\mu$m) may be 
dominated by a cold, externally heated clump, the $L_{bol}$ value 
most closely describing the protostar is likely 37 \Lsun.
\item Models that include a protostar surrounded by a disk and envelope 
with outflow cavities fit the SED well. These models yield different best-fit 
parameters depending on which photometry values are adopted and 
which model assumptions are made. Assuming a single protostar with 
an infalling envelope, we estimate that the envelope density is relatively high
($\rho_1$ $\sim$ 10$^{-13}$ $-$ 10$^{-12}$ g cm$^{-3}$ 
or $\rho_{1000}$ $\sim$ 3 $\times$ 10$^{-18}$ $-$ 
3 $\times$ 10$^{-17}$ g cm$^{-3}$), both for models with
polynomial-shaped and streamline-shaped cavities. 
\item The SED can also be fit by combining a protostellar model that 
considers fluxes between 8 and 70 $\mu$m and a clump of externally 
heated dust that fits the longer-wavelength emission. In this model the 
luminosity is dominated by the clump, and the total luminosity of the
protostar alone amounts to $\sim$ 15-25 \Lsun\ (with corresponding
$L_{bol}$ values of 12-14 \Lsun). The envelope density is still
high ($\rho_1$ close to 10$^{-13}$ or $\rho_{1000}$ close to
3 $\times$ 10$^{-18}$ g cm$^{-3}$), suggesting an early
evolutionary state for the protostar (Stage 0). Given the significant 
contribution of the molecular clump to the long-wavelength emission, 
the protostar is probably best described by this model.
\item We find that the position of OMC-2 FIR 4 measured in our
IRAC 4.5 $\mu$m image is offset with respect to the position measured
at 8-70 $\mu$m, but matches that of the radio continuum source
detected at 3.6 cm by \citet{reipurth99}. Both can be interpreted
as emission from shocked gas in an outflow. Furthermore, there is
evidence in favor of an outflow from far-IR spectra \citep{manoj13,
kama13} in the form of velocity profiles, temperatures, and densities 
derived from CO lines, although they may contain a contribution from
an outflow driven by the nearby protostar OMC-2 FIR~3. These data 
support the idea that FIR 4 is indeed a protostar, driving a compact
outflow. In addition, the centimeter flux is consistent with that 
observed in outflows from other protostars with luminosities
$<$ 100 \Lsun\ \citep{anglada95}.
\item Using fluxes measured in a 20\arcsec\ aperture centered on 
the clump position (i.e., the position of the peak flux at $\lambda$
$\geq$ 160 $\mu$m) and applying a modified blackbody fit, we
estimate a temperature of 22 K and a mass of 27 \Msun\ for the
clump. This clump could form more protostars, and OMC-2 FIR 4, 
which lies near its edge, might be the first one formed, but is probably
still growing in mass and luminosity. Thus, we agree with the suggestion
of \citet{shimajiri08} and \citet{lopez13} that the molecular core
of OMC-2 FIR 4 likely fragmented, with one of these fragments
currently containing a protostar. However, we find that the data is 
best explained by a $<$ 100 \Lsun\ protostar and not an
intermediate-mass, luminous ($\sim$ 1000 \Lsun) young star as 
proposed by \citet{crimier09} and \citet{lopez13}. Although the
protostar currently has a modest luminosity, the final stellar mass 
it will obtain is difficult to predict considering that it is embedded in 
a core with a total mass of 27 \Msun.  
\end{itemize}

Only long-wavelength observations at high spatial resolution, such as 
the VLA and ALMA can provide, will allow us to better understand  
this object. In particular, mapping the dust continuum and the outflows 
at resolutions $\lesssim$ 1\arcsec\ will constrain the envelope 
structure, including the properties of the cavity and inclination angle. 
This in turn will settle the question about this object's luminosity. 
Overall, OMC-2 FIR 4 will further our understanding of the star 
formation process in complex environments such as OMC 2.

\acknowledgments
This work is based on observations made with the {\it Spitzer Space Telescope}, 
which is operated by the Jet Propulsion Laboratory (JPL), California Institute of 
Technology (Caltech), under a contract with NASA; it is also based on
observations made with the {\it Herschel Space Observatory}, a European Space
Agency Cornerstone Mission with significant participation by NASA. 
The {\it Herschel} spacecraft was designed, built, tested, and launched under 
a contract to ESA managed by the Herschel/Planck Project team by an industrial 
consortium under the overall responsibility of the prime contractor Thales Alenia Space 
(Cannes), and including Astrium (Friedrichshafen) responsible for the payload module 
and for system testing at spacecraft level, Thales Alenia Space (Turin) responsible 
for the service module, and Astrium (Toulouse) responsible for the telescope, with 
in excess of a hundred subcontractors.
We also include data from the Atacama Pathfinder Experiment, a collaboration 
between the Max-Planck Institut f\"ur Radioastronomie, the European Southern 
Observatory, and the Onsala Space Observatory. 
Support for this work was provided by NASA through awards issued by JPL/Caltech. 
The work of A.\ M.\ S.\  was supported by the Deutsche Forschungsgemeinschaft 
priority  program 1573 (``Physics of the Interstellar Medium'').
M.\ O.\ acknowledges support from MICINN (Spain) AYA2008-06189-C03-01 and
AYA2011-30228-C03-01 grants (co-funded with FEDER funds).

\appendix

\begin{figure}[h]
\centering
\includegraphics[scale=0.65, angle=90]{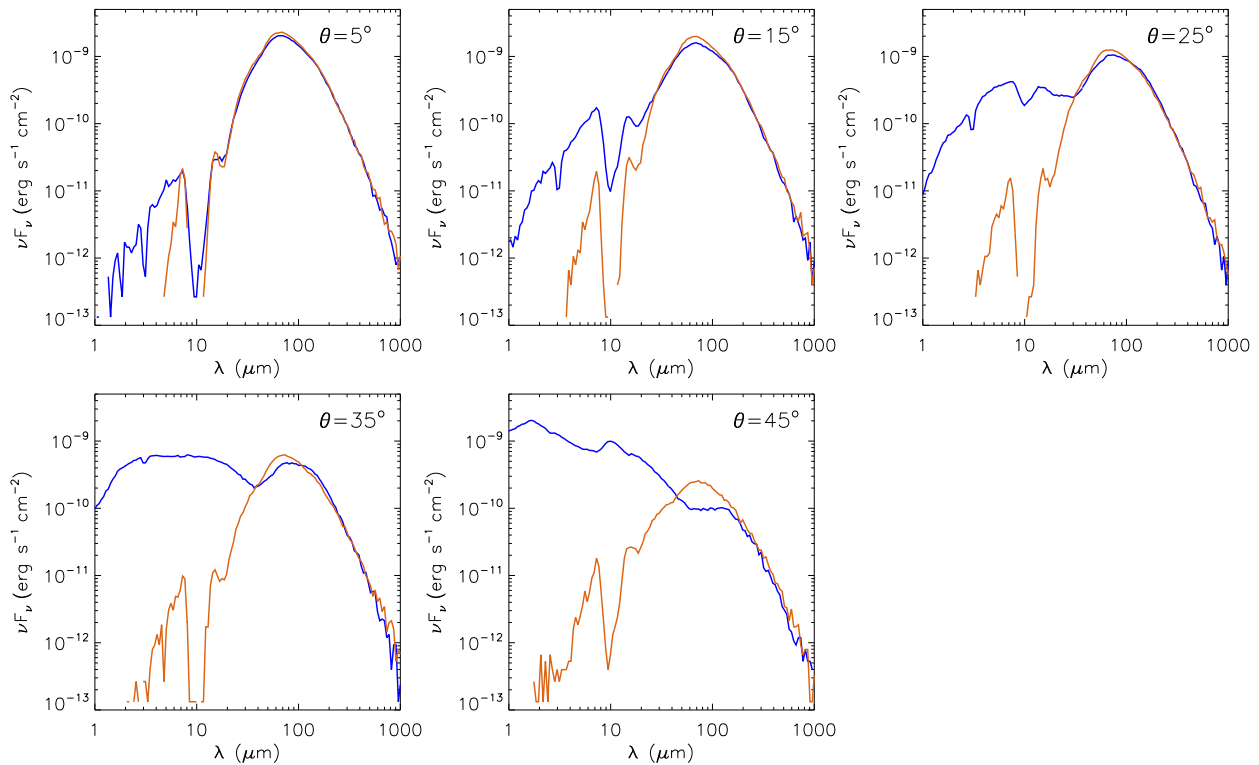}
\caption{Model 4 from section \ref{mod}, calculated with the same parameters
except for the cavity: in the five different panels, the same model is shown with
five different cavity opening angles (see label inside each panel), and in each panel 
for two different cavity shapes ({\it blue}: polynomial-shaped cavity with exponent 
1.5; {\it orange}: streamline-shaped cavity). Note that Model 4 has an inclination
angle of 49\degr, and the best fit to HOPS 108 has $\theta$=5\degr.
\label{Model4_cavity_effects1}}
\end{figure}

\begin{figure}[!]
\centering
\includegraphics[scale=0.65, angle=90]{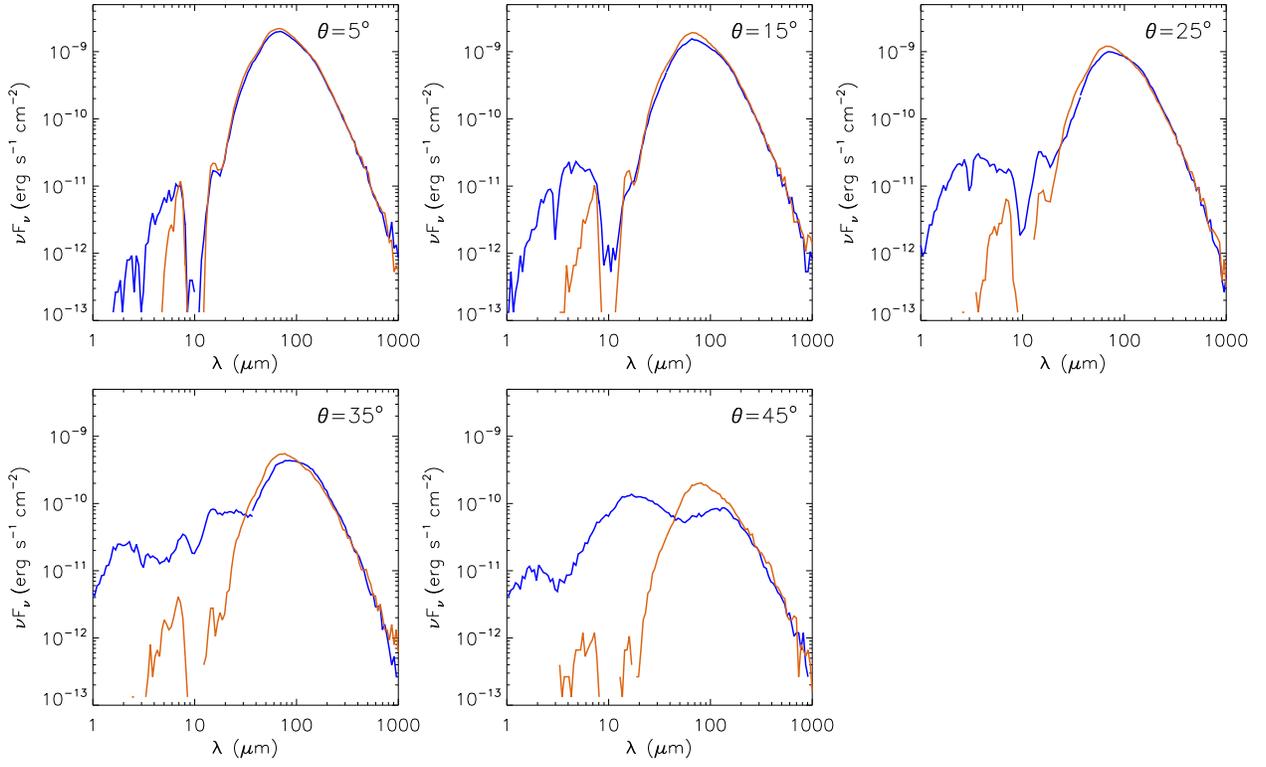}
\caption{Same as in Figure \ref{Model4_cavity_effects1}, but the models are
shown for an inclination angle of 70\degr. 
\label{Model4_cavity_effects2}}
\end{figure}

Models assuming different cavity shapes (polynomial vs.\ streamline)
can result in widely different model parameters. Here we explore the 
effect of the cavity shape on the SED. 
In Figures \ref{Model4_cavity_effects1} and \ref{Model4_cavity_effects2}
we show Model 4 from section \ref{mod} with different cavity opening
angles and at two inclination angles (49\degr\ in Figure \ref{Model4_cavity_effects1} 
and 70\degr\ in Figure \ref{Model4_cavity_effects2}; all other model
parameters are unchanged). 

At both inclination angles, the differences between models with a streamline-shaped 
cavity and those with a polynomial-shaped cavity become pronounced when 
the cavity opening angle is fairly large, $\gtrsim$ 25\degr. Then the 
polynomial-shaped cavity, which evacuates more material, allows more 
shorter-wavelength photons to reach the observer; the silicate absorption 
feature and the mid-IR SED slope become shallower. At an inclination angle 
of 49\degr, there is a noticeable difference between the two types of cavity 
already at $\theta$=15\degr. As the cavity opening angle reaches 
45\degr, the stellar and disk emission are unobscured with a polynomial-shaped
cavity, while a streamline-shaped cavity still leaves sufficient envelope
dust along the line of sight to cause a deep silicate absorption feature
and a steeply rising SED in the mid-IR. 
Only the models with $\theta$=5\degr\ agree well irrespective of
cavity shape. This is also reflected in our modeling results from section
\ref{mod}, where the best-fit model parameters of Model 4 (which
assumed a polynomial-shaped cavity and had $\theta$=5\degr) are 
very similar to those of Model 8 (which assumed a streamline-shaped
cavity and also had $\theta$=5\degr).

\end{document}